\documentclass{aa}  
\usepackage{CJKutf8}
\usepackage{graphicx}
\usepackage[breaklinks,colorlinks,citecolor=blue,linkcolor=blue]{hyperref}
\usepackage{txfonts}
\usepackage{lscape}
\usepackage{epstopdf}
\usepackage{CJK}
\usepackage{float}
\usepackage[absolute]{textpos}
\usepackage{amsmath}
\usepackage{mathrsfs}

\usepackage{arydshln}

\begin{document} 
\title{Ram pressure shaping high-velocity cloud droplets}
\subtitle{FAST HI observations of HVC AC-III and theoretical interpretation}
   \author{Xunchuan Liu \begin{CJK}{UTF8}{gbsn}(刘训川)\end{CJK}
          \thanks{liuxunchuan001@gmail.com}
          } 

   \institute{
   Leiden Observatory, Leiden University, P.O. Box 9513, 2300RA
Leiden, The Netherlands 
              }
              
\date{Received xxx xx}
\abstract{
FAST \ion{H}{i} observations reveal unprecedented internal structures within the high-velocity cloud (HVC) AC-III, which consists of several coherent subclumps (D1--D6). These subclumps exhibit nearly constant line widths of $\sim 20\,\mathrm{km\,s^{-1}}$ across a velocity ($\varv_{\rm LSR}$) range of $-220$ to $-180\,\mathrm{km\,s^{-1}}$. They display a parabolic morphology consistent with ram-pressure-confined droplets whose heads are oriented toward the Galactic plane. A steady-state analytic model can reproduce both the morphologies and the exponential density profiles of these droplets. The density at the droplet heads reaches $\sim 2\,\mathrm{cm^{-3}}$, implying an ambient gas density of $\sim 10^{-3}\,\mathrm{cm^{-3}}$, which matches the conditions of the Galactic warm ionized medium (WIM). Deviations from axial symmetry, alongside rich dynamic patterns (ridges, rings, and cavities), indicate complex internal dynamics within AC-III. These patterns may be caused by fluid loops induced by interactions between neighboring droplets. Additionally, we find an intermediate-velocity component ($-150$ to $-100\,\mathrm{km\,s^{-1}}$) showing a distinct shell-like morphology that envelops the head of AC-III, suggesting a non-contact interaction possibly mediated by the WIM. Overall, we suggest that AC-III is actively entering the Galactic WIM layer and being sculpted by external drag into a droplet-like morphology, providing an ideal laboratory for investigating the evolution of halo gas infalling toward the Galactic plane.
}

\keywords{ISM: kinematics and dynamics — ISM: structure — ISM: individual objects (HVC AC-III) — Galaxy: halo — radio lines: ISM (HI)}

 \maketitle
 

\section{Introduction} \label{sec_intro}
High-velocity clouds (HVCs) are \ion{H}{I} structures moving at velocities inconsistent with Galactic rotation \citep{1991A&A...250..499W,2018MNRAS.474..289W}.
Absorption-line studies against halo stars behind a limited sample of HVCs have detected associated ionized gas, indicating that they are multi-phase systems \citep[e.g.,][]{1999Natur.400..138V,2018MNRAS.473.5514T,2025ApJ...988..251V}.
On average, HVCs are thought to reside in pressure equilibrium with the surrounding hot halo \citep[e.g.,][]{2006A&A...455..481K}.
HVCs typically have distance brackets of one to ten kiloparsecs (kpc) above the Galactic plane \citep[e.g.,][]{1999Natur.400..138V,2008ApJ...672..298W,2008ApJ...684..364T,2016ApJ...828L..20P}, locating them at the interface of the extended hot halo (spanning several to tens of kpc) and the stratified warm ionized medium \citep[WIM, with a scale height of $\sim1$~kpc;][]{2008PASA...25..184G}.
This ionized gas may arise both from the ionization balance intrinsic to the HVC itself and from the accumulation of ambient material during interactions with the hot halo and the Galactic WIM layer \citep[e.g.,][]{1999ASPC..166..180S,2011ApJ...739...30K}.
The predominantly negative velocities of HVCs \citep{1991A&A...250..499W,2008ApJ...672..298W,2018MNRAS.474..289W} indicate that they may trace the infall of low-metallicity gas from the circumgalactic medium onto the Galactic disk \citep{1980ApJ...236..577B,2004A&A...419..527D,2007ApJ...670L.113W,2016ApJ...816L..11F,2023ApJ...944...65C}.
Therefore, HVCs are widely recognized as key components of the Galactic gas cycle, playing a crucial role in feeding the host galaxy with fresh material, sustaining star formation, and mediating the exchange of gas between the disk and the halo \citep{1980ApJ...236..577B,2008A&ARv..15..189S}.
However, there is still no consensus on their specific origins or evolutionary pathways.

Another important characteristic of HVCs is that, owing to the improved resolution of modern single-dish telescopes and interferometric arrays, their internal structures have been revealed to be far more complex than previously anticipated.
High-resolution observations show that some HVCs exhibit filamentary morphologies, often appearing as elongated or strip-like structures aligned with the large-scale features of infalling streams \citep{2008ApJ...679L..21L,2017ApJ...834..126B,2021AAS...23752709B,2024AAS...24340224H,2025MNRAS.536.3507H}.
Recent FAST observations of the very-high-velocity cloud (VHVC) G165, a subclass of HVCs with $|V_{\rm LSR}| \gtrsim 200~\mathrm{km~s^{-1}}$, reveal intricate internal structures consisting of intertwined filaments that form a complex, network-like pattern produced by turbulent compression, the detailed origin of which remains unclear \citep{2025NatAs...9.1366L}.
This morphological complexity points to rich internal processes that may arise from the interplay among turbulence, thermal instability, and environmental interactions \citep[e.g.,][]{2026arXiv260700872L}.
There are also indications that extended HVCs possess stripped substructures induced by shear-driven disturbances from the surrounding medium \citep[e.g.,][]{2021AAS...23752709B,2025NatAs...9.1366L}; however, the exact causal relationship remains ambiguous due to alternative explanations, and such features are notoriously difficult to investigate quantitatively. 
Due to resolution limitations, such well-studied cases are usually restricted to extended HVCs, which typically reside deep in the halo at distances exceeding $5$~kpc.

It is reasonable to expect that the infalling behavior of HVCs, combined with their interactions with the diverse ambient environment, provides a framework for their evolutionary pathways, shaping their global morphology, internal turbulence, dynamic structures, and multi-phase transitions. To explore this scenario, it is important to examine the characteristic dynamic patterns induced by interactions between HVCs and the surrounding gas. 
A few HVCs, such as the Smith Cloud, are thought to be undergoing tidal stripping 
after transiting the Galactic disk \citep[e.g.,][]{2009ApJ...707.1642N,2018MNRAS.473.5514T}. 
A larger population of HVCs is expected to endure milder interactions with the 
inner halo of our Galaxy, resulting in observable morphological modifications. 
Most HVCs, however, do not display morphologies clearly shaped by pressure interactions \citep[e.g.,][]{2018MNRAS.474..289W}. This absence can be attributed to the high temperature ($\sim 10^6~\mathrm{K}$) of the hot halo \citep{1976ApJ...205..762S,1980ApJ...236..577B,2024CmPhy...7..286B}, within which HVCs move largely unaffected, experiencing subsonic motion relative to the ambient gas. 
This picture may change dramatically when HVCs enter the potential cold component of the hot halo \citep[e.g.,][]{2024CmPhy...7..286B}, or more likely, the much colder ($10^4~\mathrm{K}$) and denser ($10^{-3}~\mathrm{cm^{-3}}$) Galactic layer of the WIM. This medium features a compact, disk-like structure compared to the hot halo and is constrained by the Galactic gravitational potential and magnetic field \citep{1991ApJ...372L..17R,2008PASA...25..184G,2009RvMP...81..969H}. In this environment, interactions are expected to strongly influence both the morphology and internal structure of HVCs. In such dense layers, we may expect to discover compact HVCs whose morphologies are predominantly shaped by these hydrodynamical interactions while being isolated from confounding environmental effects, thereby enabling us to study interaction-induced dynamics in detail and gain insights into HVC evolution.

We examined the HVCs revealed by the all-sky \ion{H}{I} survey HI4PI \citep{2016A&A...594A.116H,2018MNRAS.474..289W} and found that AC-III exhibits a head-tail morphology shaped by ram pressure. AC-III is located at Galactic coordinates $(l, b) = (190\degr, -30\degr)$ with a central radial velocity of $\varv_{\rm LSR} \sim -200~\mathrm{km~s^{-1}}$.\footnote{The source is located near the Galactic anti-center, where the Galactic-frame velocity approximately equals the Local Standard of Rest velocity $\varv_{\rm LSR}$.} The cloud head has a compact angular size of $\sim 2\degr$ and is oriented toward the Galactic plane, suggesting that the system is intrinsically approaching the Galactic disk at high velocity rather than its apparent motion merely resulting from a projection effect. AC-III therefore has the potential to provide an excellent laboratory for investigating dynamic interactions with the ambient medium; however, the limited angular resolution of HI4PI (beam size $\sim 16.2\arcmin$) makes it difficult to resolve its internal structures in sufficient detail to place effective physical constraints on these processes. This limitation motivated us to observe the AC-III complex using FAST, which provides a significantly superior resolution of $\sim 3\arcmin$.

In this work, we present the droplet-like substructures of AC-III resolved by FAST observations and provide a theoretical interpretation of its morphology and internal density profiles within the framework of ram-pressure–induced droplets and their localized inter-droplet interactions. The paper is structured as follows. Sect.~\ref{sec_fastobs} describes the FAST observations and data processing. Sect.~\ref{sec_results} presents the observed morphology and kinematics of AC-III. In Sect.~\ref{sec_steadysolv}, we introduce the analytical steady-state droplet model and examine the geometric projection effects on the observed morphology and column density distributions. Sect.~\ref{sec_aciiifit} presents the physical properties of the droplets and ambient gas inferred from droplet fitting. Sect.~\ref{sec_dynaeq} investigates how local interactions cause departures from symmetric droplet morphologies, producing internal linear ridges, rings, and cavities inspired by our observations. Sect.~\ref{sec_discuss} evaluates the possible interactions between the HVC and the WIM, as well as the separate lower-velocity \ion{H}{I} gas layers in the AC-III region. Finally, Sect.~\ref{sec_summary} summarizes our main findings and conclusions.

\section{Observations and data reduction}\label{sec_fastobs}
\subsection{AC-III}
\begin{figure}[!t]
    \centering
    \includegraphics[width=0.99\linewidth]{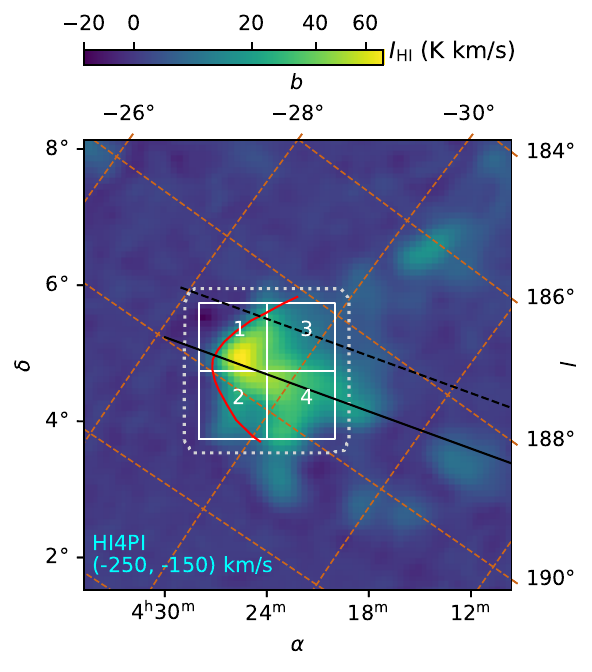}
\caption{HI4PI \citep{2016A&A...594A.116H} column density (moment-0) map of AC-III, integrated over the velocity range of $-250$ to $-150$\,$\mathrm{km\,s^{-1}}$. White squares indicate the four on-the-fly (OTF) fields observed with FAST (see Sect.~\ref{sec_fastobs}). The dotted white line outlines the full FAST spatial coverage, including the noisy edge margins (see the upper-left panel of Fig.~\ref{noisetime}). The red curve traces the global trajectory of the ram-pressure-induced parabolic front, identified from a joint analysis of the HI4PI and FAST data cubes. Solid and dashed black lines indicate the trajectories of the two position--velocity cuts, labeled PV1 and PV2, respectively, which are analyzed in Sect.~\ref{sec_IVC}. The orange grid marks the Galactic longitude ($l$) and latitude ($b$) coordinates.}
    \label{fig:HI4PImom0}
\end{figure}

At the Galactic anti-center ($l \sim 180\degr$), the HVC populations are dominated by Chain~A, the AC group, and the Cohen Stream \citep{2018MNRAS.474..289W}. Chain~A, the AC shell, AC-I/II, and the Cohen Stream are all extended, elongated structures spanning several tens of degrees. In contrast, AC-III is a more compact complex with an angular cross-section of a few degrees (Fig.~\ref{fig:HI4PImom0}), situated between AC-I/II and the Cohen Stream. While the distance to AC-III remains poorly constrained, neighboring anti-center structures have well-established distance bounds. Specifically, \citet{1996ApJS..104...81T} suggested that the AC shell lies within $\sim 6.5$~kpc of the Sun, while \citet{2008ApJ...672..298W} estimated a distance bracket of $5.0\text{--}11.7$~kpc for the Cohen Stream. Furthermore, Chain~A has been measured to reside within $10$~kpc \citep{1999Natur.400..138V}. These independent constraints imply that the HVCs in the anti-center sky are relatively nearby and likely associated with the local Galactic disk–halo interface. For the analytical models in this work, we adopt a representative distance of $5$~kpc for AC-III.

\begin{figure*}[!t]
    \centering
    \includegraphics[width=0.99\linewidth]{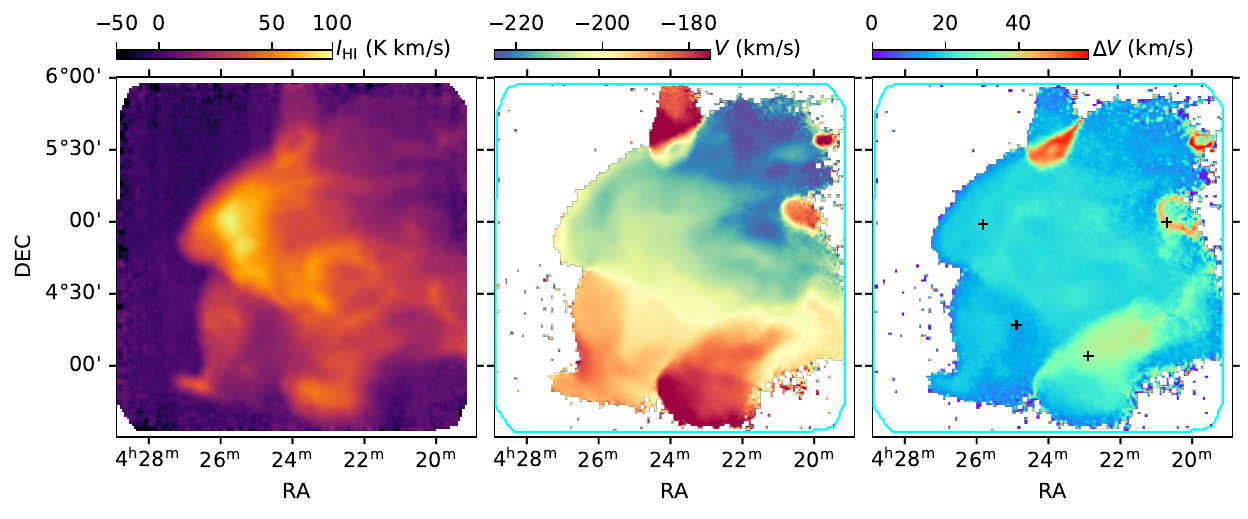}
\caption{
Moment maps of the FAST \ion{H}{I} emission for AC-III, integrated over the velocity range $-250$ to $-160~\mathrm{km\,s^{-1}}$. 
The moment~0 map ($M_0$ or $I_{\rm HI}$, left) is displayed without a clipping threshold, whereas the mean velocity ($M_1$ or $V$, middle) and line-width ($\Delta V$, calculated via Eq.~\ref{eqM2DV}, right) maps are masked to include only channels exceeding $3.5\sigma$ (Sect. \ref{sec_results}). 
The cyan contour in the middle and right panels outlines the full FAST observational coverage (see Fig.~\ref{fig:HI4PImom0}). 
Individual spectra corresponding to the four black crosses marked in the right panel, labeled p1--p4 from left to right, are presented in Fig.~\ref{fig_speexample}.
\label{fig_mommaps}
}
\end{figure*}

\subsection{FAST observations}\label{sec_fastobsdetail}
We conducted FAST \citep{2011IJMPD..20..989N} observations of the high-velocity cloud AC-III (FAST project ID: PT2025\_0043, PI: Xunchuan Liu) between 2025 August 12 and 15. The data were collected in the on-the-fly (OTF) mapping mode utilizing the $L$-band 19-beam receiver array. The low declination of the target ($\delta \sim 5^\circ$) allowed all scans to be performed within the telescope's optimal zenith angle range ($\theta_{\rm ZA} < 26.4^\circ$), where the illuminated aperture reaches its full 300~m diameter \citep{2019SCPMA..6259502J}.

The $L$-band front end provides full-polarization measurements, recording the auto-correlations (XX and YY) alongside the real and imaginary components of the cross-correlations (XY). In this work, we focus exclusively on the Stokes~$I$ intensity, obtained by averaging the two orthogonal auto-correlations. The telescope features a typical full-width at half-maximum (FWHM) beam size of $\sim 3'$ at 1.4~GHz, a system temperature below 24~K, and a beam efficiency of 0.85 for extended sources \citep{2019SCPMA..6259502J,2022A&A...658A.140L}, operating with a tracking accuracy of 0.2\arcmin. The digital spectral backend records 1024k channels over a 500~MHz bandwidth (1.0--1.5~GHz), yielding a frequency resolution of $\delta f \sim 0.48$~kHz per channel. This corresponds to a velocity resolution of $\sim 0.1~\mathrm{km\,s^{-1}}$ at the 1.42~GHz \ion{H}{I} line rest frequency. Raw spectra are sampled every $dt_{\rm sample} = 240 \times t_{\rm period,0} \sim 0.5033$~s, where $t_{\rm period,0} = 2^{21}$~ns $\sim 0.00209$~s is the elemental time interval of an individual Fast Fourier Transform (FFT). Precise knowledge of this sampling cadence is critical, as the data reduction pipeline relies on an exact time-to-coordinate mapping to project channel indices onto sky positions \citep{2025NatAs...9.1366L}.

AC-III was covered with four square fields (Fig.~\ref{fig:HI4PImom0}), each measuring $1\degr \times 1\degr$. To satisfy the Nyquist sampling criteria between scan rows, the feed cabin was rotated clockwise by 53.4\degr\ \citep[for details on the scan pattern, see][]{2025NatAs...9.1366L}. The mapping was performed by scanning along declination at a speed of 30\arcsec~s$^{-1}$, with successive rows separated in right ascension by 1.7\arcmin. This redundant sampling improves sensitivity and suppresses striping artifacts caused by differences in individual beam responses. To cover each field, 36 rows were scanned. Including a 54~s turnaround time ($t_{\rm return}$) for each row, each square field required 6,210~s of telescope time (not including a standard 10-minute overhead). In total, 8 hours of telescope time were used to cover the four fields.

Flux calibration was anchored using the low-noise injection mode of a temperature-stabilized noise diode ($T_{\rm cal} \sim 1.1~\rm K$). The modulated noise signal was injected every 2~s (corresponding precisely to intervals of $4\,dt_{\rm sample}$) with symmetrical on/off states. Each individual injected noise pulse lasted for $2\,dt_{\rm sample}$, strictly synchronized with the hardware sampling cadence of the spectral backend.

\subsection{Data reduction}
The data reduction process, which includes gridding, Doppler correction, flux calibration, spectral baseline subtraction for each field, and merging the data cubes into a single cube, follows the procedures described in \citet{2025NatAs...9.1366L}. During the observations, strong radio frequency interference (RFI) occasionally occurred in short intervals of about 1~minute. The affected data points were removed according to the methods detailed in Appendix~\ref{sec_rfi}.

\section{Observational results of HVC AC-III}\label{sec_results}
We define the $n\text{-th}$ order moment ($M_n$) of a spectrum $T(\varv)$ as:
\begin{equation}
    M_n = \frac{\int (\varv - \varv_0)^n \, T(\varv)\, d\varv}{\int T(\varv)\, d\varv},
\end{equation}
where $\varv_0 = 0$ for $n=1$, and $\varv_0 = M_1$ for $n>1$. Here, $M_0$ and $M_1$ correspond to the integrated intensity ($I$) and the mean velocity ($V$) of the spectrum, respectively. The $M_2$-derived FWHM line width is then given by:
\begin{equation}
    \Delta V = \sqrt{8 \ln 2 \, M_2}. \label{eqM2DV}
\end{equation}
Fig.~\ref{fig_mommaps} shows the moment maps ($I_{\rm HI}$, $V$, and $\Delta V$) of the \ion{H}{I} emission in AC-III observed with FAST, integrated over the velocity range $-250$ to $-160$~$\mathrm{km\,s^{-1}}$.

\begin{figure}[t]
    \centering
    \includegraphics[width=0.99\linewidth]{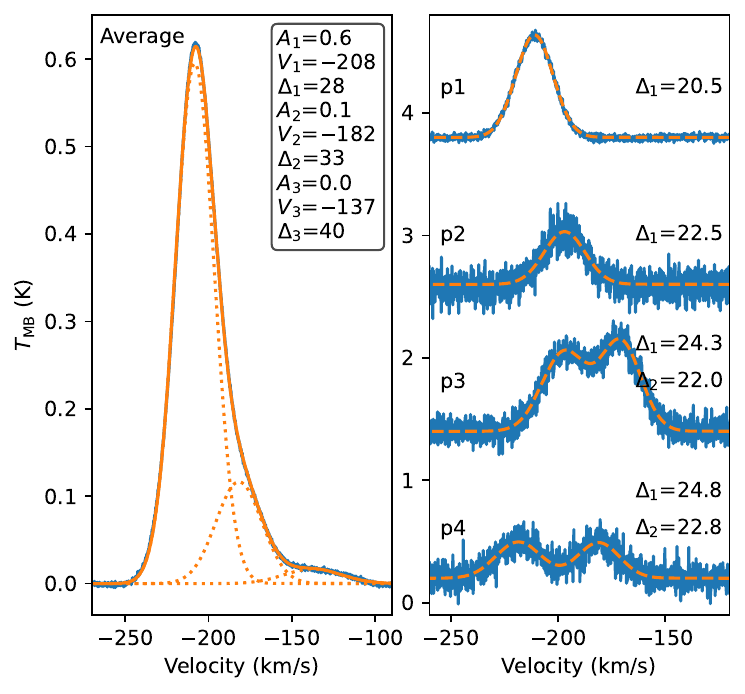}
\caption{
Left: Mean \ion{H}{I} spectrum (blue) averaged over the FAST-covered region of AC-III, plotted together with a three-component Gaussian fit (orange). The resulting Gaussian parameters are listed in the upper right corner, including the peak brightness temperature ($T_{\rm peak}$), the central velocity ($V$), and the FWHM line width ($\Delta$) for each individual component. Right: Same as the left panel but for the localized spectra extracted at positions p1--p4 (marked in the right panel of Fig.~\ref{fig_mommaps}). The fitted FWHM line widths for each position are explicitly labeled. For visual clarity, the intensity of the spectrum at p1 has been scaled down by a factor of five.
\label{fig_speexample}}
\end{figure}

The average spectrum of AC-III within the FAST-covered region can be decomposed into three Gaussian components (Fig.~\ref{fig_speexample}). The velocity range used in Fig.~\ref{fig_mommaps} covers the two major velocity components spanning $\varv \sim -250$ to $-160$~$\mathrm{km\,s^{-1}}$, while a third, relatively weak component appears intermittently in the range $\varv \sim -150$ to $-100$~$\mathrm{km\,s^{-1}}$. Below, we focus on the two major components. The nature of the weak component, including whether it represents a distinct intermediate-velocity cloud (IVC) or is dynamically associated with AC-III, is discussed further in Sect.~\ref{sec_IVC}.

\subsection{Subclumps with constant line widths} \label{sec_tlw} 

The observed \ion{H}{I} spectra exhibit well-defined Gaussian profiles with typical line widths of $\sim 20~\mathrm{km~s^{-1}}$. At positions where the $M_2$-derived line width appears large, the spectra can be decomposed into two distinct Gaussian components, each retaining a line width of $\Delta V \sim 20~\mathrm{km~s^{-1}}$ (see Fig.~\ref{fig_speexample} for examples). Even in pixels where the $M_2$ method yields narrower widths (such as position p2 in the right panel of Fig.~\ref{fig_mommaps}), multi-component Gaussian fitting still recovers a characteristic width of $\Delta V \sim 20~\mathrm{km~s^{-1}}$. This consistency demonstrates that the reduced $M_2$ values in those regions are an artifact of low-intensity profiles where the faint Gaussian wings drop below the masking threshold, rather than an intrinsic physical narrowing of the gas.

A line width of $\Delta V \sim 20~\mathrm{km~s^{-1}}$ corresponds to a Doppler temperature of:
\begin{equation}
    T_{\rm Doppler} = \frac{m_{\rm H} \, \Delta V^2}{8 \ln 2 \, k_{\rm B}} \sim 8700\ \mathrm{K},
\end{equation}
where $m_{\rm H}$ is the mass of a hydrogen atom and $k_{\rm B}$ is the Boltzmann constant. This value is significantly higher than the lower limit of the spin temperature ($\sim 700\ \mathrm{K}$) estimated against the background continuum source (Appendix~\ref{sec_foretemp}). While this gap prevents us from explicitly disentangling the exact fractions contributed by thermal versus turbulent broadening, a maximum kinetic temperature below $10\,000\ \mathrm{K}$ strongly indicates that the gas is dominated by the warm neutral medium (WNM). This line width is also highly typical of classical WNM profiles \citep{2003ApJ...586.1067H,2023ARA&A..61...19M,2025NatAs...9.1366L}, further confirming that this phase dominates the neutral gas component of AC-III.

These spectral behaviors indicate that AC-III is structurally composed of multiple coherent subclumps, each characterized by a nearly constant Gaussian line width of $\sim 20~\mathrm{km~s^{-1}}$. Based on their distinct spatial boundaries detailed in Sect.~\ref{sec_droplets}, we hereafter refer to these individual subclumps as "droplets." Under this structural framework, the broader global velocity distribution of AC-III, which spans from $-220$ to $-180~\mathrm{km~s^{-1}}$ (middle panel of Fig.~\ref{fig_mommaps}), is not driven by large internal thermal gradients, but is instead produced by the relative bulk motions among these individual droplets. Consequently, AC-III can be modeled as a coherent parcel of subsonic or transonic WNM, where the bulk of the system's kinetic energy is stored in these clump-to-clump motions.

\subsection{Density and mass} \label{sect_meandensity}
When the \ion{H}{I} emission is optically thin, the conversion between the integrated intensity and the \ion{H}{I} column density is given by \citep[e.g.,][]{2003ApJ...585..823L,2023ARA&A..61...19M}:  
\begin{equation}
    N(\ion{H}{I}) = 1.82 \times 10^{18} \,
    \frac{I_{\rm HI}}{\rm K\,km\,s^{-1}} \ {\rm cm}^{-2}.
    \label{eq_HI_N}
\end{equation}
For a single Gaussian profile, the integrated line intensity scales as:
\begin{equation}
    I_{\mathrm{HI}} = \sqrt{\frac{\pi}{4\ln 2}}\, T_{\mathrm{peak}}\, \Delta V,  
\end{equation}
where $T_{\rm peak}$ is the peak brightness temperature. Adopting the characteristic line width of $\Delta V = 20~\mathrm{km\,s^{-1}}$ derived in Sect.~\ref{sec_tlw}, the column density expression simplifies to:
\begin{equation}
    N(\ion{H}{I}) \approx 3.9 \times 10^{19} 
    \left(\frac{T_{\rm peak}}{\rm K}\right) \ {\rm cm}^{-2}.
\end{equation}

The peak brightness temperature, averaged over the typical $1^\circ$ angular scale of the individual subclumps, is $\sim 1$~K. The corresponding mean volume density can be estimated by assuming a spherical geometry:
\begin{equation}
    n \sim \frac{N(\ion{H}{I})}{d \times 1^\circ} \sim 0.2\, d_5^{-1} \ {\rm cm^{-3}},
    \label{eq_aven}
\end{equation}
where $d$ is the distance and $d_5 = d / 5$~kpc. This value is highly typical of the WNM phase \citep{2003ApJ...586.1067H,2025NatAs...9.1366L}. The total integrated flux over the FAST-covered region is 75~K~km~s$^{-1}$~deg$^2$, corresponding to a total \ion{H}{I} mass of:
\begin{equation}
    M_{\rm HI} \sim 8.5\times10^3\, d_5^2\ M_\odot.
\end{equation}
Assuming a 1D velocity dispersion of $\sigma_\nu = 10~\mathrm{km\,s^{-1}}$ and an angular radius of $1^\circ$, the required virial mass is:
\begin{equation}
    M_{\rm vir} = \frac{R\sigma_\nu^2}{G} \sim 2\times10^6\ d_5\ M_\odot.
\end{equation}
Because any added contributions from bulk clump-to-clump motions would only increase the effective velocity dispersion, the true $M_{\rm vir}$ is even higher. Consequently, the condition $M_{\rm vir} \gg M_{\rm HI}$ holds across all plausible distance brackets, confirming that the cloud complex is gravitationally unbound.

The discrepancy of nearly two orders of magnitude between the two quantities cannot be reconciled by un-modeled baryonic mass components. The derived mean density of \(\sim 0.2~\mathrm{cm^{-3}}\) indicates that the hydrogen gas is predominantly in the atomic phase \citep[e.g.,][]{2003ApJ...587..278W}, meaning that \(M_{\rm HI}\) represents the bulk of the baryonic matter. Therefore, self-gravity plays a negligible role in the cloud’s dynamical evolution, and alternative external support mechanisms must be considered.

\begin{figure}
    \centering
    \includegraphics[width=0.9\linewidth]{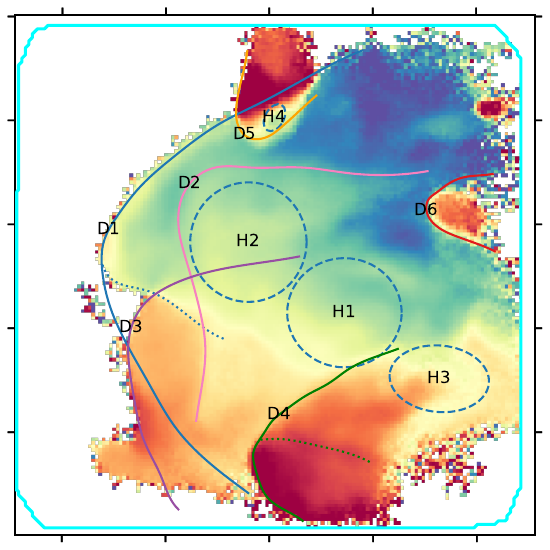}
    \includegraphics[width=0.9\linewidth]{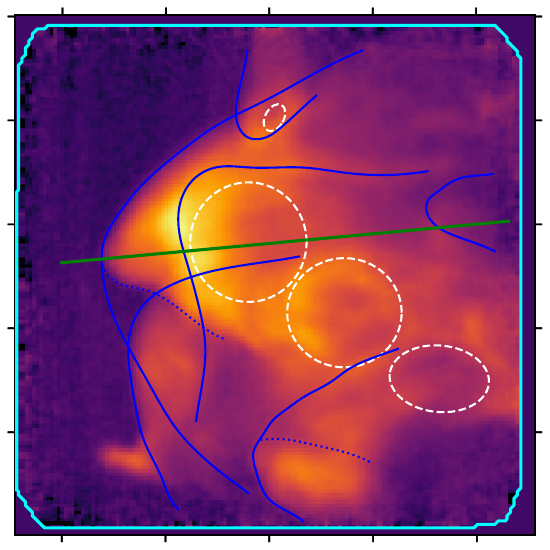}
\caption{Moment~1 (upper panel) and moment~0 (lower panel) maps, presented in an identical layout to Fig.~\ref{fig_mommaps}, but overlaid with annotations for individual droplets and rings. Solid curves highlight the boundaries of individual droplet-like features (D1--D6), while ellipses indicate the ring-like emission enhancements enveloping the internal cavities. The dotted lines mark the distorted boundaries (Sect. \ref{sec_droplets}). The color scales and coordinate ticks match those in the corresponding panels of Fig.~\ref{fig_mommaps} and are omitted for brevity. The blue curve tracking the head of D1 in the upper panel corresponds to the large-scale parabolic front marked by the red curve in Fig.~\ref{fig:HI4PImom0}. The solid green line in the lower panel denotes the projected axis of the distorted D1, along which the density profile is shown in Fig.~\ref{fig:fitmainaxis}.}
    \label{fig:fronts}
\end{figure}

\begin{figure*}[!t]
\centering
\includegraphics[width=0.99\linewidth]{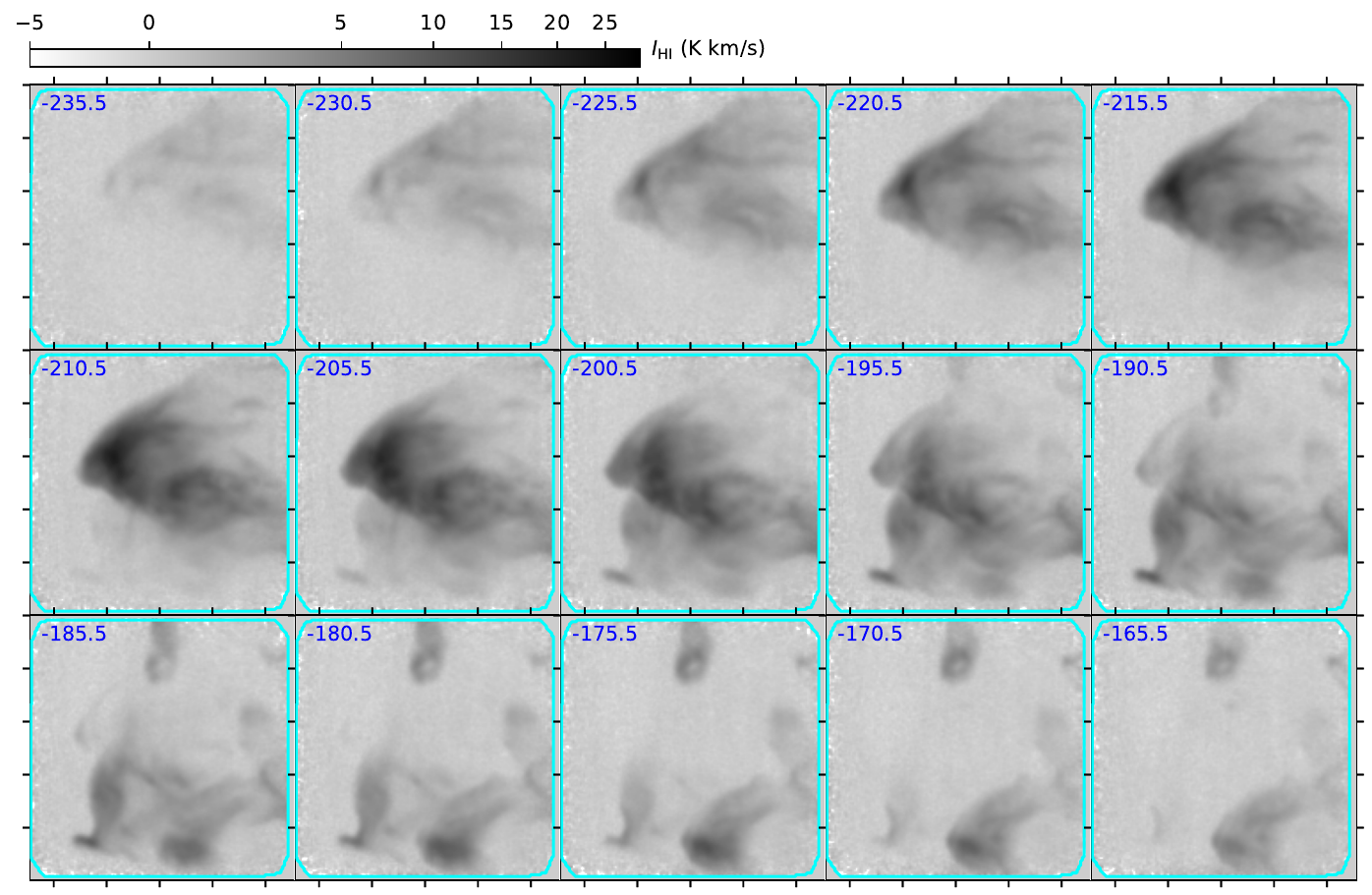} 
\caption{Channel maps of the \ion{H}{I} emission for AC-III observed with FAST. Each panel is integrated over a velocity interval of 5~$\mathrm{km\,s^{-1}}$, with consecutive panels offset by a uniform velocity step. The central velocity of each interval (in $\mathrm{km\,s^{-1}}$) is indicated in the upper-left corner of each panel. All panels share a common color bar. The coordinate ticks are identical to those in Fig.~\ref{fig_mommaps}, and the cyan contour outlines the boundary covered by the FAST observations. \label{fig_channelmaps}
}
\end{figure*}

\subsection{Subclumps as droplets} \label{sec_droplets}

The moment~0 maps can be decomposed into a number of subclumps whose spatial boundaries trace approximately parabolic profiles. These structures are even more clearly delineated in the moment~1 maps, where their parabolic droplet heads are especially prominent. The column density typically peaks within the head region and decreases rapidly toward the broader tail. We interpret these symmetric profiles as droplets formed under external ram-pressure confinement (Sect.~\ref{sec_steadysolv}). In total, six such droplets are identified and labeled D1 through D6 (Fig.~\ref{fig:fronts}). Throughout this work, unless otherwise noted, references to cardinal directions are given in Galactic coordinates, where north points toward the Galactic plane and west points in the direction of Galactic rotation.

To constrain the global motion of AC-III, we examined the geometric orientations of the droplet heads (Fig.~\ref{fig:fronts}), which show modest variations but are overall systematically biased toward the Galactic plane. The predominantly negative line-of-sight velocities of AC-III cannot be explained by standard Galactic rotation due to its anti-center location at $l \sim 190^\circ$. Assuming a circular Galactic rotation speed of $220\,\mathrm{km\,s^{-1}}$ and a simple flat rotation curve \citep{1998gaas.book.....B}, the maximum possible line-of-sight velocity contributed by Galactic rotation is:
\begin{equation}
v_{\mathrm{rot, LOS}} = 220\,\sin(190^\circ - 180^\circ) \sim 38\,\mathrm{km\,s^{-1}}, \label{eq_Grot}
\end{equation}
which is far smaller than the observed bulk velocity of AC-III ($\sim -200\,\mathrm{km\,s^{-1}}$).
Moreover, Eq.~\ref{eq_Grot} predicts a positive (receding) velocity, in direct contrast to the observed negative (approaching) value. Thus, the large observed negative velocity must be dominated by the cloud's intrinsic peculiar motion toward the Galactic disk. Combined with this rapid infall, the parabolic profiles of the droplets indicate that they are being actively sculpted by ram pressure as the complex plunges into the Galaxy. In this scenario, the external drag naturally compresses the neutral gas into the forward droplet head region (Sect.~\ref{sect_steady_ram}).

Further examination reveals that the projected boundaries of several droplets overlap along the line of sight (Fig.~\ref{fig:fronts}). The most prominent structure is droplet D1, which spans an angular size of $\sim 2^\circ$ with a droplet head pointing toward the Galactic north. Droplets D3 and D4 are fainter but comparable in scale ($\sim 1^\circ$) and line up along the eastern boundary of D1. In contrast, D5 and D6 are more compact ($\sim 0.2^\circ$ across) and are situated near the western boundary and the trailing tail of D1, respectively. Spectrally, components D3--D6 exhibit central velocities $\gtrsim 20~\mathrm{km~s^{-1}}$ higher (i.e., redder) than D1 (see also the channel map in Fig.~\ref{fig_channelmaps}). Droplet D2 shows a distinct head embedded within the projected area of D1, with its eastern tail extending toward D3. While the $M_2$-derived line widths are generally enhanced at these apparent intersection zones (Fig.~\ref{fig_mommaps}), such spectral broadening does not by itself prove a physical interaction, since simple line-of-sight projection can mimic this effect. Additional spatial and kinematic evidence is needed to determine whether these droplets are dynamically coupled.

The morphology of D1 reveals clear deviations from an ideal, ram-pressure-shaped steady state. Its droplet head is asymmetrical, with the eastern\footnote{Directions are referenced in Galactic coordinates throughout (see Fig.~\ref{fig:HI4PImom0}).} boundary compressed inward compared to the more rounded western side. In Fig.~\ref{fig:fronts}, we reconstruct the undistorted eastern boundary (solid blue curve in the upper panel) from the observed distorted boundary (dashed blue curve) by extrapolating from the unaffected western side based on: (1) symmetry with respect to the unperturbed western boundary; (2) the large-scale morphology of AC-III tracked by HI4PI (Fig.~\ref{fig:HI4PImom0}); and (3) alignment with the outer boundaries of D2, D3, and D4. As the dominant component of AC-III, D1 acts as the primary shield resisting the external flow, largely governing the overall configuration of the system. Interestingly, the western boundaries of D3 and D4 intrude into the projected boundary of D1. While the overlapping gas profiles of D3 are difficult to disentangle, the structural distortion of D4 is clearly visible as an asymmetric velocity gradient in the moment~1 map.

Inhomogeneities in the ambient medium may cause asymmetric morphologies in individual droplets. However, localized external density fluctuations alone cannot account for why the distorted sides differ so randomly among neighboring droplets. The physical intrusion of neighboring boundaries, combined with these systematic departures from axial symmetry, suggests that active inter-droplet interactions are occurring. The spatial arrangement of these neighboring clumps may, in turn, perturb the ambient flow, leading to local variations in the effective ram pressure (Sect.~\ref{sec_pattern}). Overall, these droplets are likely shaped by a combination of a global ram-pressure background and localized, close-range inter-droplet compression, creating a dynamically complex environment that drives the internal velocity patterns observed within the individual droplets (Sect.~\ref{sec_dpatterns}).

\subsection{Dynamic patterns: strips, rings, and holes in droplets}
\label{sec_dpatterns}
The superior angular resolution of FAST, compared to previous full-sky surveys such as HI4PI, allows us to resolve not only the parabolic morphology of the individual droplets but also their detailed internal density profiles. These distinct features, including linear ridges, rings, and cavities, provide new insights into the localized dynamical processes operating within the droplets.

Unlike VHVC G165, which harbors a highly complex, web-like filamentary network, AC-III exhibits relatively simpler internal sub-structures. These features appear as linear ridges preferentially oriented along the middle axes of the droplets (the central symmetry axes resolved in the projected plane of observation), suggesting that the internal density profiles are structurally modulated by the background external ram pressure. However, an ideal droplet in a steady state would remain featureless and devoid of such internal substructures. Their presence confirms that internal gas dynamics must also play a primary role, which is entirely consistent with the observed velocity gradients directed both parallel and perpendicular to these middle symmetry axes (Fig.~\ref{fig:fronts}).

In addition to these linear ridges, another notable structural feature is the presence of rings and cavities. Along the eastern boundary of D1, three distinct cavities (H1–H3) spanning diameters of $\sim 0.5^\circ$ are visible, each outlined by localized emission enhancements that form coherent, ring-like structures (Fig.~\ref{fig:fronts}). Among them, H2 represents the most ideal case, exhibiting a highly symmetric circular profile. These three rings are roughly aligned along the middle axis of D2, suggesting that D2 may not represent an external, projecting droplet like D3 or D4, but rather a dynamic structure intrinsically embedded within the main body of D1. This interpretation is further supported by the absence of any substantial line-width enhancement associated with the D2 region. A smaller cavity (H4), with an angular diameter of approximately $8\arcmin$, is also detected within the overlap zone between D1 and D5. Taken together, these findings indicate that rings and cavities represent a common internal pattern arising from anisotropic hydrodynamical interactions between neighboring droplets.

In the following two sections, we develop a theoretical framework to explain how these droplet morphologies are established and how they depart from ideal forms. In Sect.~\ref{sec_steadysolv}, we derive the analytical steady-state solution for a droplet shaped by uniform ram pressure, assuming a static density distribution. In Sect.~\ref{sec_dynaeq}, we demonstrate that highly anisotropic external pressure fields cannot be balanced by a steady-state density distribution, and must instead induce a dynamic velocity field within the droplets. Under this scenario, the internal velocity can remain time-independent, allowing the resulting density structures to form a stable, long-lived pattern that persists over extended evolutionary timescales.

\section{Steady-state pressure-confined gas} \label{sec_steadysolv}

\begin{figure}
    \centering
    \includegraphics[width=0.9\linewidth]{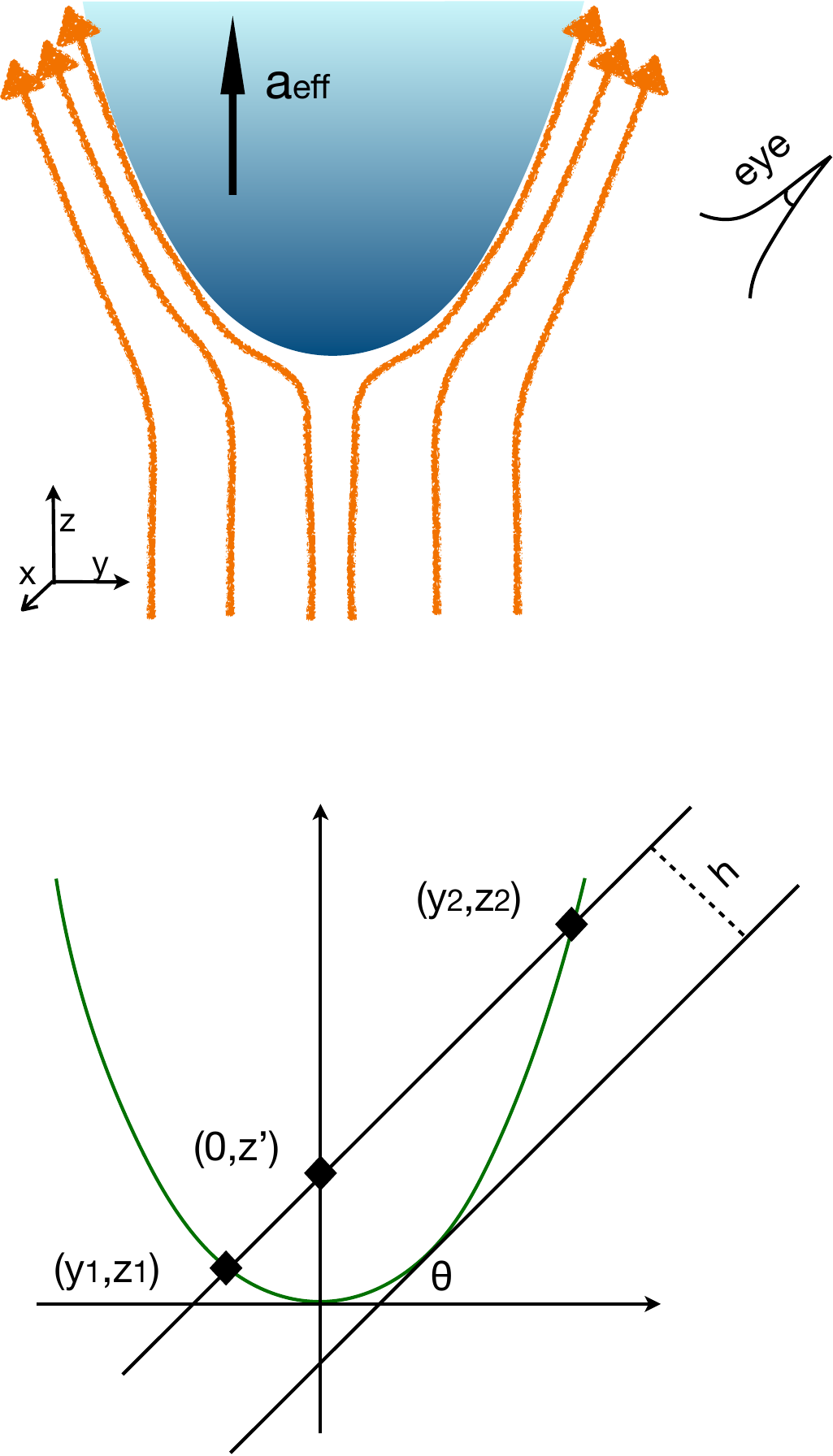}
\caption{
Upper panel: Schematic illustration of an isothermal gas droplet in a steady state (blue, with color depth representing the local volume density), confined by external ram pressure from the surrounding flows (orange arrows; see Sect.~\ref{sect_steady_ram}). Lower panel: Geometric projection of the droplet at an inclination angle $\theta$ relative to the observer (Sect.~\ref{sec_proj_steady}), where $\theta = 0\degr$ corresponds to a line of sight perpendicular to the droplet's symmetry axis.
}
\label{fig:placeholder}
\end{figure}

\begin{figure*}
    \centering
    \includegraphics[width=0.95\linewidth]{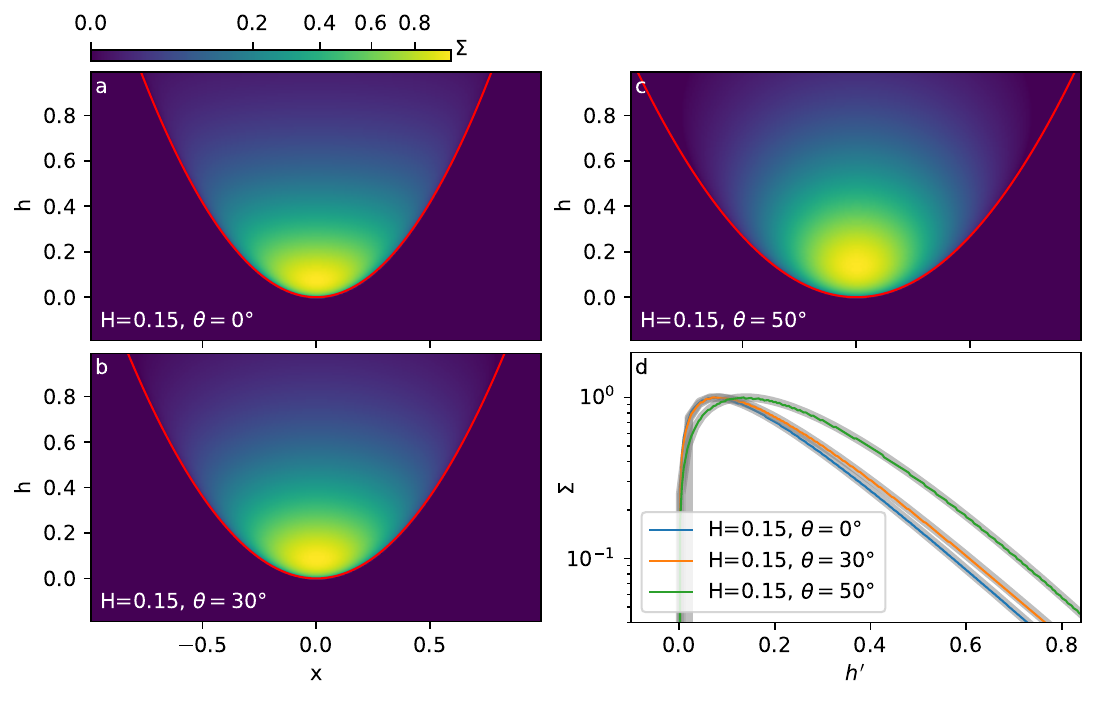}
\caption{
Panels a--c: Projective column densities ($\Sigma$, calculated via Eq.~\ref{eq_Int_parabola}) derived from a 3D numerical construction based on the analytical steady-state solution described in Sect.~\ref{sect_steady_ram}, displayed at different inclination angles (labeled in white). The column density maps are normalized to a peak value of unity. The solid red curve in each panel outlines the exact analytical boundary given by Eq.~\ref{eq_projboundary}. Panel d: Extracted column densities along the middle symmetry axis ($h'$) for each inclination angle (colored curves) compared directly with the closed-form analytical solution derived in Sect.~\ref{sec_proj_steady} (thick gray curve).
\label{fig:modelSigma} }
\end{figure*}

In this section, we present a framework describing how the geometry of the droplets is
related to some fundamental parameters, including the droplet density
and the strength of the ram pressure endured by the droplets.
The physical parameters obtained from these relations for AC-III are presented in Sect.~\ref{sec_aciiifit} and further discussed in Sect.~\ref{sec_discuss}.
Here, we assume that the gas droplets maintain a steady shape with constant internal support (see Sect.~\ref{sec_tlw}).
A simple extension of this analytical framework to include internal dynamics and droplet interactions is presented in Sect.~\ref{sec_dynaeq}, where some of the foundational equations used in this section (such as the Bernoulli invariant) can be straightforwardly applied.

We only consider systems with a time-independent density field,
\begin{equation}
    \frac{\partial \rho}{\partial t} = 0, \label{eq_constpho}
\end{equation} 
since we assume that these droplets are confined by external pressure constraints to form a stable morphology. Within this framework, a configuration completely devoid of internal motion,
\begin{equation}
    \mathbf{V} = 0,
\end{equation}
is defined as a steady state, whereas a configuration characterized by a time-independent velocity field,
\begin{equation}
    \frac{\partial \mathbf{V}}{\partial t} = 0,
\end{equation}
is defined as an equilibrium state. Considering the rich dynamic patterns observed within the droplets (Sect. \ref{sec_dpatterns}), an equilibrium state provides a more complete description. However, as a first-order approximation, we adopt a steady-state assumption throughout this section.

\subsection{Density distribution under Bernoulli equilibrium}\label{eq_rhober}

Since the droplets we observed have nearly constant line widths (Sect.~\ref{sec_tlw}), we assume the gas follows an ideal isothermal equation of state:
\begin{equation}
    p = T \rho,
\end{equation}
with the rescaled temperature parameter $T$ defined as:
\begin{equation}
    T = \frac{k_{\rm B} T_{\rm phys}}{\mu m_{\rm H}} = c_{\rm s}^2,
\end{equation}
where $\mu$ is the mean molecular weight, $T_{\rm phys}$ is the physical temperature, $m_{\rm H}$ is the hydrogen atom mass, $k_{\rm B}$ is the Boltzmann constant, and $c_{\rm s}$ represents the isothermal sound speed. When macroscopic turbulent support is considered (Sect.~\ref{sect_turb}), the total effective temperature,
\begin{equation}
    T_{\rm eff} = c_{\rm s}^2 + \sigma_{\rm turb}^2,
\end{equation}
should be used, where $\sigma_{\rm turb}$ is the 1D velocity dispersion of the small-scale turbulence.

Under an equilibrium state, any fluid loop or streamline obeys the compressible, isothermal Bernoulli invariant \citep[see, e.g.,][or Eq.~\ref{eq_global_Bc} of this work]{2002apa..book.....F}:
\begin{equation} \label{eq_theBc}
    \frac{1}{2} |\mathbf{V}|^2 + T \ln \rho + \Phi = B_c,
\end{equation}
where $\Phi$ represents the gravitational potential field and $B_c$ is the Bernoulli constant, which remains invariant along a specific fluid loop but can vary between different loops. For the steady-state assumption ($\mathbf{V} = 0$) adopted here, the Bernoulli invariant becomes globally uniform ($B_c = \mathrm{const}$ everywhere; see Eq.~\ref{eq_constBc} in Sect.~\ref{sec_decomp}). Consequently, Eq.~\ref{eq_theBc} reduces directly to:
\begin{equation}
    T \ln \rho + \Phi = 0,
\end{equation}
where $\Phi$ is a conservative scalar potential. Here, by redefining the zero-point of $\Phi$, the Bernoulli constant has been absorbed into the potential field. This relation represents a direct balance between the potential $\Phi$ and the environmental pressure $p_S$, which in turn determines the boundary potential $\Phi_S$.

If we further assume that the potential $\Phi$ arises solely from external sources and neglect self-gravity within the droplet, the potential field satisfies the Laplace equation:
\begin{equation}
    \nabla^2 \Phi \big|_{\Omega} = 0.
\end{equation}
Consequently, $\Phi$ becomes a harmonic function inside the region $\Omega$ and is uniquely determined by its boundary values. Under this condition, the corresponding steady-state density field simplifies to:
\begin{equation}
    \rho = \exp(-\Phi/T), \label{eq_rhoexp}
\end{equation}
reproducing the standard Boltzmann distribution from equilibrium statistical mechanics.


\subsection{Droplet geometry under ram-pressure deceleration} \label{sec_steadyundera}\label{sect_steady_ram}

We assume that the gravitational field and the deceleration rate induced by ram pressure remain constant over the dynamical timescale of the droplets. The external gravitational field is uniform across the droplets, which are small compared to the scale over which the background field varies (as estimated from the distance to the cloud), allowing tidal forces to be safely neglected. The potential $\Phi$ can then be modeled as a linear scalar potential corresponding to a uniform acceleration, representing the combined effect of a uniform gravitational field $\mathbf{g}$ and an imposed global deceleration $\mathbf{a}$. The potential is expressed as:
\begin{equation}
    \Phi(\mathbf{r}) = \mathbf{a}_{\rm eff} \cdot \mathbf{r}, \quad \mathbf{a}_{\rm eff} = \mathbf{a} - \mathbf{g}, \label{eq_aeff}
\end{equation}
where $\mathbf{r}$ is the position vector originating at the droplet head. We assume that $\mathbf{a}$ and $\mathbf{g}$ are in the same direction, then the droplet is expected to exhibit axial symmetry along the direction of $\mathbf{a}_{\rm eff}$. 
In Sects.~\ref{sec_scaleheight} and \ref{sect_headcur}, we show that in 3D the density scale height along the middle symmetry axis of the droplet and the curvature of the droplet head are tightly correlated, both being uniquely determined by $\mathbf{a}$ and the internal temperature. In Sect.~\ref{sec_proj_steady}, we examine how 2D projection affects these properties.

\subsubsection{Density scale height} \label{sec_scaleheight}

Eqs.~\ref{eq_rhoexp} and \ref{eq_aeff} imply that the density exhibits an exponential gradient along the symmetry axis of a droplet:
\begin{equation}
    \rho(\mathbf{r}) = \rho_0 \exp\Big(-\frac{\mathbf{a}_{\rm eff} \cdot \mathbf{r}}{T}\Big)= \rho_0\exp\Big(-\frac{a_{\rm eff} h}{T}\Big), \label{eq_rho_uniforma}
\end{equation}
where $h$ is the distance from the droplet head measured along the direction of the middle symmetry axis (corresponding to the $z$-axis in Fig.~\ref{fig:placeholder}), $a_{\rm eff}=\| \mathbf{a}_{\rm eff} \|$, and $\rho_0$ is the density at the droplet head (also referred to as the head peak density). This exponential distribution naturally features a characteristic scale height defined as:
\begin{equation}
    H = \frac{T}{a_{\rm eff}}. \label{eq_T_a}
\end{equation}
To maintain this internal configuration, the confinement boundary pressure  cannot be arbitrary; it must structurally follow the same exponential spatial profile dictated by Eq.~\ref{eq_rho_uniforma}.

\begin{figure}[!t]
    \centering
    \includegraphics[width=0.99\linewidth]{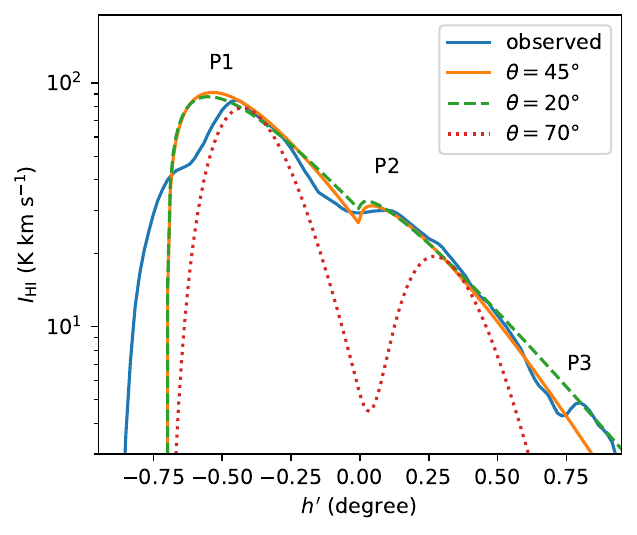}
    \includegraphics[width=0.99\linewidth]{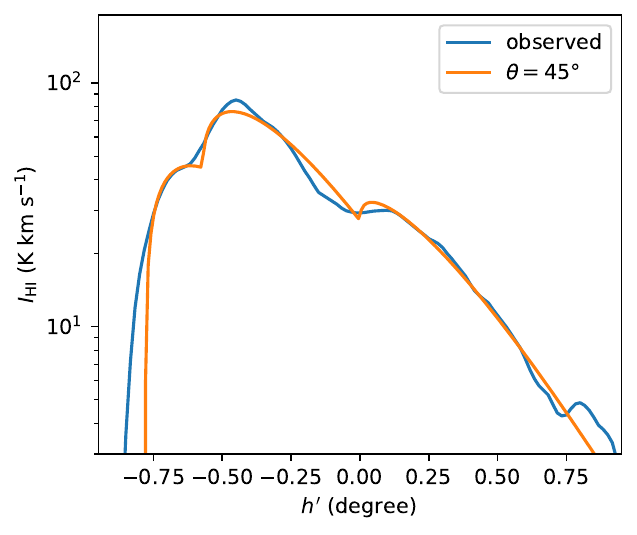}
\caption{Upper: The solid blue curve represents the projected \ion{H}{i} column density profile (derived from the integrated intensity) extracted along the solid green line in the lower panel of Fig.~\ref{fig:fronts}, which is adopted as the projected axis ($h'$, Eq.~\ref{eq_projh}) of the distorted D1 droplet (Sect.~\ref{sec_droplets}). The orange, green, and red curves display the corresponding multi-component model fits (Sect.~\ref{sec_dropfit}) adopting different inclination angles ($\theta$, Sect.~\ref{sect_steady_ram}). The first peak (P1) corresponds to the tips of D1 ($h'\sim-0.75\degr$) and D2 ($h'\sim-0.5\degr$), which are not separately fitted due to their close separation of $\sim$0.25\degr. The dip between P1 and P2 marks the hole of H2, while P2 receives contributions from the edges of H2 and potentially H1. Because P2 does not explicitly correspond to a single droplet component, we assume that under global deceleration, this density enhancement obeys a similar exponential decay along the axis (Sect.~\ref{eq_rhober}). The peak P3 at $h'=0.8\degr$ is contributed by D6, which is omitted from the fit because it is too weak.
Lower: Same as the upper panel, but with $\theta$ fixed at 45\degr, and the peaks of D1 and D2 are fitted separately.}
    \label{fig:fitmainaxis}
\end{figure}

The effective cross-sectional area of a droplet depends on the geometry of its droplet head (see Sect.~\ref{sect_headcur}); however, the deceleration rate can be well approximated using its physical cross-sectional area, $A$. To physically ground this ram-pressure acceleration, consider a gas cloud exposed to an external flow with a density $\rho_{\rm ext}$ and a relative velocity $\mathbf{v}_{\rm ram}$ intercepting this area. The bulk hydrodynamic drag force exerted by the ram pressure can be expressed as:
\begin{equation}
    F_{\rm ram} \sim \rho_{\rm ext} \|\mathbf{v}_{\rm ram}\|^2 A, \label{eqFram}
\end{equation}
which acts against the cloud's inertia. Defining the average column density (mass per unit area) of the gas as:
\begin{equation}
    \Sigma = \frac{M}{A},
\end{equation}
the resulting net macroscopic deceleration experienced by the droplet becomes:
\begin{equation}
    a_{\rm eff} \sim \frac{F_{\rm ram}}{M} = \frac{\rho_{\rm ext} \|\mathbf{v}_{\rm ram}\|^2}{\Sigma}. \label{eq_aesti}
\end{equation}
The characteristic density scale height can then be estimated by combining Eqs.~\ref{eq_T_a} and \ref{eq_aesti}:
\begin{equation}
    H \sim \frac{T}{a_{\rm eff}} = \frac{T \, \Sigma}{\rho_{\rm ext} \|\mathbf{v}_{\rm ram}\|^2}.
    \label{eq_scaleh}
\end{equation}
In Sect.~\ref{dis_halo}, we will apply this scaling relation to place empirical constraints on the density of the external medium surrounding AC-III.

\subsubsection{Head curvature} \label{sect_headcur}
The shape of the droplet boundary is determined by a local pressure balance between the internal boundary pressure, governed by Eq.~\ref{eq_rho_uniforma}, and the force exerted by the external flow. We assume the unperturbed external flow moves along the positive $z$-direction (Fig.~\ref{fig:placeholder}). To utilize the axial symmetry of the system, we adopt cylindrical coordinates and define the distance from the symmetry axis as:
\begin{equation}
    r = \sqrt{x^2 + y^2},
\end{equation}
where $x$ and $y$ represent the two orthogonal spatial directions perpendicular to the middle symmetry axis ($z$-axis; see Fig.~\ref{fig:placeholder}). The front boundary layer of the droplet head can be described by a smooth function, $z(r)$. The outward-pointing unit normal vector $\mathbf{n}$ forms an inclination angle $\alpha$ with respect to the $z$-axis. 

Balancing the pressures acting perpendicular to the front boundary layer yields the localized relation:
\begin{equation}
    p_0 \cos^2\!\alpha = p_0 \exp\!\left(-\frac{z}{H}\right), \label{eq_surfacebalance}
\end{equation}
where $p_0 = \rho_{\rm ext} \, \|\mathbf{v}_{\rm ram}\|^2$ represents the peak ram pressure of the external flow at the head peak position. In this balance, the first factor of $\cos\alpha$ accounts for the normal component of the incoming flow velocity, while the second factor accounts for the geometric area projection ratio between the intercepted flow column and the surface element. 

Using the geometric identity:
\begin{equation}
    \cos^2\!\alpha = \frac{1}{1+\left(\frac{dz}{dr}\right)^2},
\end{equation}
Eq.~\ref{eq_surfacebalance} becomes a first-order differential equation for the surface profile. Solving this relation under the standard boundary condition $z(0)=0$ provides the exact mathematical surface shape:
\begin{equation}
    z(r) = -\,2H \ln\!\left[\cos\!\left(\frac{r}{2H}\right)\right], 
    \qquad r < \pi H.
\end{equation}
Close to the central symmetry axis ($r \ll H$), a standard Taylor expansion simplifies this exact profile into a standard parabola:
\begin{equation}
    z(r) \sim \frac{r^2}{4H}. \label{eq_simplestzr}
\end{equation}
This result implies that close to the droplet head, a perfect droplet (free from extra physical effects that break axial symmetry or cause localized dynamic patterns) exhibits the shape of an axisymmetric paraboloid with a focal length equal to the density scale height $H$.

\begin{figure}[!t]
    \centering
    \includegraphics[width=0.9\linewidth]{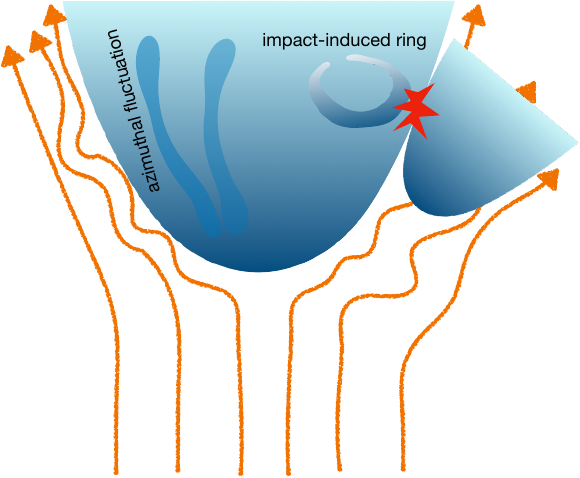}
\caption{Schematic illustration of the dynamic structural patterns within ram-pressure-confined droplets. The rotating ring acts to redistribute internal momentum and balance hydrodynamical interactions between neighboring droplets. Localized linear density ridges along the boundary layer and the middle symmetry axis are induced by external turbulence within the high-Reynolds-number surrounding flows (see Sect.~\ref{sec_pattern}).}
    \label{fig:dynamicpattern}
\end{figure}

\subsection{Projection of a paraboloid droplet} \label{sec_proj_steady}
There is a close relationship between the scale height and the curvature of the 
droplet head (Eq.~\ref{eq_simplestzr}). If the droplet is observed from a line of 
sight perpendicular to its axis (corresponding to a view angle $\theta=0\degr$ in Fig.~\ref{fig:placeholder}), 
$H$ can be determined by fitting the curvature of the droplet head, from which the volume density 
is easily deduced. In practice, however, projection effects for $|\theta|>0\degr$ 
complicate the analysis. Consequently, we must simultaneously fit the droplet 
boundary and the column density along the middle symmetry axis in the projected plane to extract both the scale height ($H$) 
and the droplet head density ($\rho_0$).

To address the projection effect, we assume an axisymmetric paraboloid droplet with a boundary 
described by:
\begin{equation}
    z = \frac{x^2+y^2}{4H}.
\end{equation}
We observe this structure within the $x=0$ plane, where the line of sight is inclined at a view angle $\theta$ relative to the $y$-axis (lower panel of Fig.~\ref{fig:placeholder}).
Consider the intersection between the parabola:
\begin{equation}
    z = \frac{y^2}{4H}
\end{equation}
and a straight line representing a tilted front:
\begin{equation}
    z = \tan\theta \, y + z',
\end{equation}
where $z'$ is the line offset. Solving:
\begin{equation}
    \frac{y^2}{4H} = \tan\theta \, y + z'
\end{equation}
yields the two intersection ordinates:
\begin{equation}
    y_{1,2} = 2H \tan\theta \mp 2 \sqrt{H^2 \tan^2\theta + H z'},
\end{equation}
with corresponding heights:
\begin{equation}
    z_{1,2} = \tan\theta \, y_{1,2} + z'.
\end{equation}
The horizontal separation is:
\begin{equation}
    \Delta y = y_2 - y_1 = 4 \sqrt{H^2 \tan^2\theta + H z'},
\end{equation}
and the line-segment length along the tilted front is:
\begin{equation}
    \Delta L = \frac{\Delta y}{\cos\theta}.
\end{equation}
The local density at ($y_1$, $z_1$) is:
\begin{equation}
    \rho_{\rm 1} = \rho_0 \exp\left(-\frac{z_1}{H}\right).
\end{equation}
Along the segment, the vertical coordinate increases as $z(s) = z_1 + s \sin\theta$, where $s$ is the distance along the segment, causing the density to decay as $\exp(-s \sin\theta / H)$. Therefore, the exact column density over the segment is given by:
\begin{equation}
    \Sigma = \rho_{\rm 1} \, \Delta L \; 
        \frac{1 - \exp\left(- \Delta L \, \sin\theta / H\right)}{\Delta L \, \sin\theta / H}. \label{eq_Int_parabola}
\end{equation}
For $\theta=0\degr$, Eq.~\ref{eq_Int_parabola} leads to the analytical expression:
\begin{equation}
    \Sigma = \rho_0 \exp(-z'/H) \times 4\sqrt{Hz'}.
\end{equation}

For $|\theta|>0\degr$, Eq.~\ref{eq_Int_parabola} has no simple analytical expression but can be solved numerically.
Real intersections exist only if:
\begin{equation}
    H^2 \tan^2\theta + H z' \ge 0,
\nobreak\end{equation}
which requires:
\begin{equation}
    z' \ge - H \tan^2\theta.
\end{equation}
The distance measured relative to the droplet head along the 2D projected symmetry axis is:
\begin{equation}
    h'= \left(z' + H\tan^2\theta\right) \cos\theta. \label{eq_projh}
\end{equation}
The $\Sigma$ maps for the same droplet projected at different view angles
are shown in panels a--c of Fig.~\ref{fig:modelSigma}. 
The profiles of $\Sigma$ (which is proportional to $I_{\rm HI}$ for optically thin objects like HVCs) along the middle symmetry axis in the projected plane as a function of $h'$ are shown in Fig.~\ref{fig:modelSigma}d.
The projected scale height ($H_{\rm vert}$), determined from the decay rate of $I_{\rm HI}$ at large values of $h'$, shows little variation across different values of $\theta$; its value is only slightly smaller for $\theta>0\degr$ than for $\theta=0\degr$ (where $H_{\rm vert}$ reduces back to $H$).

Geometrically, the projection of a quadric surface of revolution always yields a quadric curve in the plane of the sky. For an axisymmetric paraboloid droplet, its projected apparent boundary is mathematically guaranteed to outline a  parabola. By finding the envelope of the lines of sight tangent to the 3D surface, the projected half-width $R$ (defined as the transverse semi-width of this 2D boundary) can be mapped directly as a function of the distance along the 2D projected symmetry axis $h'$:
\begin{equation}
    R = \sqrt{4H^2 \tan^2\theta + 4Hz'} 
      = 2\sqrt{\frac{Hh'}{\cos\theta}}. \label{eq_projboundary}
\end{equation}
Thus, the 2D projected apparent boundary retains a strictly parabolic form. Because the projection geometry effectively stretches this profile across the transverse plane, the apparent focal parameters are scaled by a factor of $1/\cos\theta$ compared to the uninclined case (Eq.~\ref{eq_simplestzr}). Consequently, fitting the transverse envelope of this projected boundary yields an apparent horizontal scale height of:
\begin{equation}
    H_{\rm hori} = H/\cos\theta. \label{eq_Hori}
\end{equation}
Similarly to $H_{\rm vert}$, $H_{\rm hori}$ also reduces back to $H$ when $\theta=0\degr$.

In practice, we can first obtain the apparent horizontal scale height $H_{\rm hori}$ by fitting the observed droplet boundary in the projected plane. For a given $H_{\rm hori}$, different inclination angles $\theta$ lead to different intrinsic scale heights $H$ (Eq.~\ref{eq_Hori}), such that the projected column density profile $\Sigma$ becomes uniquely determined as a function of $\theta$, with its overall amplitude scaled by the front entry density $\rho_1$ (Eq.~\ref{eq_Int_parabola}). By comparing this modeled $\Sigma$ profile with the observed column density profile (or the profile of $I_{\rm HI}$) extracted along the middle symmetry axis, we can break the projection degeneracy to simultaneously determine $\rho_1$, $\theta$, and consequently the intrinsic scale height $H$.

\section{Droplet fitting of AC-III} \label{sec_aciiifit}

\subsection{Qualitative evaluation}\label{sec_roughdropfit}
The external ambient density can be reasonably constrained using Eq.~(\ref{eq_scaleh}) without relying on detailed structural modeling. The column density, $\Sigma$, along the central symmetry axis is expected to be of the same order of magnitude as along other lines of sight. We adopt a representative value of $9.1 \times 10^{19}\,\mathrm{cm^{-2}}$, derived from a typical \ion{H}{i} integrated intensity of $50\,\mathrm{K\,km\,s^{-1}}$ (see Fig.~\ref{fig_mommaps}). To reconstruct the full 3D space motion, we de-project the observed line-of-sight velocity of $\varv_{\rm GSR} \sim \varv_{\rm LSR} \sim -200\,\mathrm{km\,s^{-1}}$ (Sect.~\ref{sec_intro}). Assuming an inclination angle of $\theta = 45^\circ$,\footnote{An inclination of $\theta = 0^\circ$ would result in a zero line-of-sight velocity component, whereas $\theta = 90^\circ$ would fail to preserve the observed droplet-like geometry. We therefore assume an intermediate value between these two extremes.} we apply a geometric correction factor of $1/\sin(45^\circ) = \sqrt{2}$ to the velocity magnitude, yielding an estimated total external flow velocity of:
\begin{equation}
    \|\mathbf{v}_{\rm ram}\| \sim \sqrt{2} \times 200 \sim 300\,\mathrm{km\,s^{-1}}.
\end{equation}
The spatial scale height of the droplet is taken as $H \sim d \times 0.25\degr$ (where $0.25\degr$ is approximately the observed droplet head size of D1), and the effective internal temperature of AC-III is estimated from the line width as:
\begin{equation}
    T_{\rm eff} \sim \left(\frac{\Delta V}{2}\right)^2, \quad \Delta V \sim 20\ \mathrm{km\,s^{-1}}.
\end{equation}
The resulting volume number density of the external medium is given by:
\begin{equation}
    n_{\rm ext} \sim \frac{T_{\rm eff}\Sigma}{H \|\mathbf{v}_{\rm ram}\|^2} \sim \frac{1.4\times 10^{-3}}{d_5}\ \mathrm{cm^{-3}}. \label{eq_esnext}
\end{equation}
This external density is low, indicating that the surrounding gas should reside in a highly ionized state \citep{2003ApJ...587..278W}, which is entirely consistent with its non-detection in deep 21-cm \ion{H}{I} emission surveys.

\subsection{Quantitative fitting}\label{sec_dropfit}
We can estimate the intrinsic 3D density scale height, $H$, and the head peak mass density, $\rho_0$ (or head peak number density, $n_{\rm head}$), of AC-III by accounting for projection effects from a non-zero inclination angle, $\theta$, following the framework described in Sect.~\ref{sec_proj_steady}. 
To extract the geometric parameters, we first apply a quadratic curve fit via Eqs.~\ref{eq_projboundary} and \ref{eq_Hori} to the droplet boundary, which was traced as the solid blue curve in the upper panel of Fig.~\ref{fig:fronts} (see Sect.~\ref{sec_droplets} for details on the boundary extraction). This yields a horizontal scale height of:
\begin{equation}
    H_{\rm hori} \sim d \times 0.3^\circ.
\end{equation} 
To mitigate line-of-sight blending from D3, D4, and H1, we extract the \ion{H}{i} column density profile of D1 along the central axis of the distorted D1 feature (marked by the solid green line in the lower panel of Fig.~\ref{fig:fronts}). The resulting profile fit derived from Eq.~\ref{eq_Int_parabola} is displayed in Fig.~\ref{fig:fitmainaxis}. However, blending (especially from H2) still affects the extraction path; consequently, a single profile given by Eq.~\ref{eq_Int_parabola} cannot fully reproduce the observations. To circumvent this, we employ a combination of two such profiles sharing the same scale height, $H$, and inclination angle, $\theta$, to fit the observed data (upper panel of Fig.~\ref{fig:fitmainaxis}). One corresponds to the combined contribution of D1 and D2, while the other marks the density enhancement from H2 and potentially H1. A three-component fit, achieved by separating the tips of D1 and D2 (lower panel of Fig.~\ref{fig:fitmainaxis}), yields a density decrease of roughly half for D1, but the combined density of D1 and D2 remains the same. Note that because the slope of the tail provides the critical constraint on the inclination, introducing a new fitting component does not provide additional constraints on the fitted parameters. Below, we base our discussion on the two-component fit (upper panel of Fig.~\ref{fig:fitmainaxis}).

This structural complexity prevents the inclination angle $\theta$ from being uniquely constrained. As shown in the upper panel of Fig.~\ref{fig:fitmainaxis}, moderate inclination angles (such as $20\degr$ or $45\degr$) can reasonably reproduce the observed column density profile, whereas values close to $90\degr$ are strictly ruled out. This result implies that AC-III cannot be moving purely along the line of sight; its velocity component perpendicular to the line of sight must be considerable.

Note that a $\theta<|b|=30\degr$ will lead to a outward motion of the system in Galactic radius direction.
We adopt a representative inclination angle of $\theta = 45\degr$. 
The density scale height can be evaluated from Eq. \ref{eq_Hori}: 
\begin{equation}
    H = H_{\rm hori}\cos(\theta) = d\times 0.21\degr. \label{eq_Hfor45}
\end{equation}
An angular size of $0.21^\circ$ is approximately four times the FAST beam size (Sect.~\ref{sec_fastobsdetail}). Convolving a structure of this angular scale with the beam will broaden the apparent angular size by less than 5\%. Beam smearing by the telescope is therefore not expected to significantly affect the fitting results.
The head peak number density fitted from Eq. \ref{eq_Int_parabola} is 
\begin{equation}
    n_{\rm head} \sim 2\,d_5^{-1}\ \mathrm{cm^{-3}},
\end{equation} 
which is approximately ten times higher than the average volume density of the cloud (see Eq.~\ref{eq_aven} in Sect.~\ref{sect_meandensity}). From Eq.~\ref{eq_scaleh}, this relates to the external medium via:
\begin{equation}
    n_{\rm ext} \sim n_{\rm head} \frac{\|\mathbf{v}_{\rm internal}\|^2}{\|\mathbf{v}_{\rm ram}\|^2}  
    \sim n_{\rm head} \frac{\|\mathbf{v}_{\rm internal}\|^2}{\|\mathbf{v}_{\rm LOS}/\sin\theta\|^2},
\end{equation}
where $\mathbf{v}_{\rm internal}$ is the local velocity dispersion at the droplet head (including both thermal and turbulent motions), adopted as $\|\mathbf{v}_{\rm internal}\| \sim \Delta V/2 \sim 10\ \mathrm{km\,s^{-1}}$, and $\|\mathbf{v}_{\rm LOS}\|$ is the observed line-of-sight velocity. An external density of $n_{\rm ext} \sim 2\times 10^{-3}\ \mathrm{cm^{-3}}$ is then obtained, confirming the result derived via the qualitative approach in Sect.~\ref{sec_roughdropfit}.



\section{Equilibrium via internal motion} \label{sec_dynaeq}
For a global equilibrium state, as opposed to an instantaneous steady state, the situation is generally more complex. When the external pressure profile deviates from a simple exponential form, maintaining equilibrium would require a correspondingly complex external potential $\Phi$, which is rarely satisfied in practice. Instead, the system typically responds by developing an anisotropic stress tensor, such as a Reynolds stress tensor, to support the resulting pressure gradients. In principle, the fully general equilibrium problem is therefore highly complicated, but it remains essential for explaining the asymmetric morphology and internal dynamic patterns that deviate from perfect droplets as described in Sect.~\ref{sec_steadysolv}.

\subsection{Vorticity-induced Bernoulli balance}\label{sec_decomp}


This subsection summarizes standard results from classical fluid dynamics for completeness. 
We consider a steady-state, compressible, isothermal ideal gas. In the rest frame of 
the droplet, the standard advective form of the momentum conservation equation (the inviscid 
Euler equation) reads 
\begin{equation}
\rho (\mathbf{V} \cdot \nabla) \mathbf{V} + \nabla p = - \rho \nabla \Phi,
\end{equation}  
where $\mathbf{V}$ is the velocity field of the external ambient gas relative to the droplet 
head, $\rho$ is the ambient gas density, $p$ is the gas pressure, 
and $\Phi$ represents the external gravitational potential.
Substituting the isothermal equation of state ($p = T \rho$) and dividing by $\rho$, 
the momentum equation can be expressed in terms of the density gradient as
\begin{equation}
(\mathbf{V} \cdot \nabla) \mathbf{V} = - T \nabla \ln \rho - \nabla \Phi.
\end{equation}
By applying a standard vector identity,\footnote{Note that 
$(\mathbf{V} \cdot \nabla) \mathbf{V} = \nabla \left( \frac{1}{2}|\mathbf{V}|^2 \right) - \mathbf{V} \times (\nabla \times \mathbf{V})$.} 
this relation can be rewritten in a form that isolates the spatial gradients:
\begin{equation}
\nabla \underbrace{\left(\frac{1}{2}|\mathbf{V}|^2 + T \ln \rho + \Phi \right)}_{B_c} 
= \mathbf{V} \times (\nabla \times \mathbf{V}), \label{eq_global_Bc}
\end{equation}
where $B_c$ denotes the Bernoulli quantity (see Eq.~\ref{eq_theBc}).

The cross-product term on the right-hand side represents the Lamb vector, which physically captures the rotational inertial force. Because this vector is strictly orthogonal to the local velocity vector ($\mathbf{V}$), taking the scalar product with $\mathbf{V}$ demonstrates that the Bernoulli quantity is invariant along any given streamline:
\begin{equation}
\mathbf{V} \cdot \nabla B_c = 0.
\end{equation}
Conversely, across different streamlines, the spatial variation of the Bernoulli quantity is governed entirely by the flow vorticity:
\begin{equation}
\nabla_\perp B_c = \mathbf{V} \times (\nabla \times \mathbf{V}).
\end{equation}
Finally, for a steady state, the gradient vanishes ($\nabla B_c = 0$), meaning that
\begin{equation}
    B_c = \mathrm{const} \quad \text{for} \quad \mathbf{V}=0. \label{eq_constBc}
\end{equation}

To illustrate this behavior under simple rotational conditions, consider a rigid-body rotation in a plane described by $\mathbf{V} = \mathbf{\Omega} \times \mathbf{r}$, where $\mathbf{\Omega}$ is a constant angular velocity. In this configuration, the vorticity field is uniform ($\nabla \times \mathbf{V} = 2\mathbf{\Omega}$), which simplifies the Lamb vector to the gradient of the kinetic energy:
\begin{equation}
\mathbf{V} \times (\nabla \times \mathbf{V}) = \nabla \left( |\mathbf{V}|^2 \right).
\end{equation}
Substituting this result back into Eq.~\ref{eq_global_Bc} balances the kinetic energy gradient on both sides of the equation. Upon cancellation, the expression simplifies directly to
\begin{equation}
T \ln \rho + \Phi - \frac{1}{2}|\mathbf{V}|^2 = \text{const}, \label{eq_rotBc}
\end{equation}
where $-\frac{1}{2}|\mathbf{V}|^2$ naturally emerges and acts as the centrifugal potential. It implies that an ordered eddy can provide 
constant pressure support.

\begin{figure}
    \centering
    \includegraphics[width=0.99\linewidth]{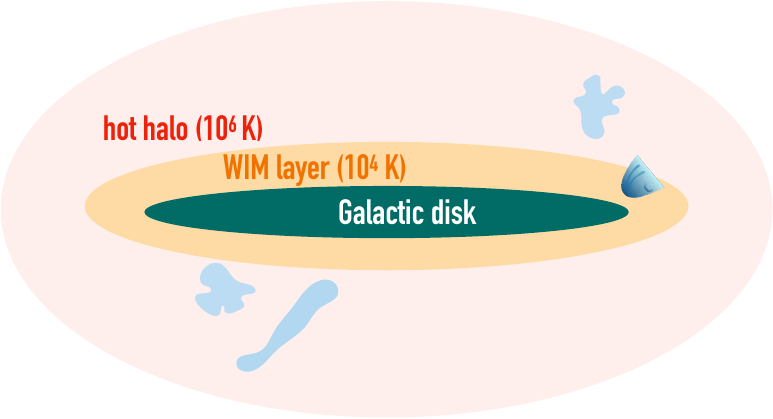}
\caption{Schematic overview of high-velocity clouds (HVCs) within the Galactic halo environment. 
AC-III (indicated in red) is actively plunging into the disk-stratified Warm Ionized Medium (WIM) layer that blankets the Galactic disk (see Sects.~\ref{dis_halo} and \ref{sec_external}). 
In contrast, the majority of classical HVC complexes (indicated in blue) float safely at larger Galactocentric distances (ranging from several to tens of kiloparsecs) within the diffuse hot halo medium (see Sects.~\ref{sec_dropfit} and \ref{dis_halo}).
\label{fig_sketch-large}}
\end{figure}

\begin{figure}
    \centering
    \includegraphics[width=0.9\linewidth]{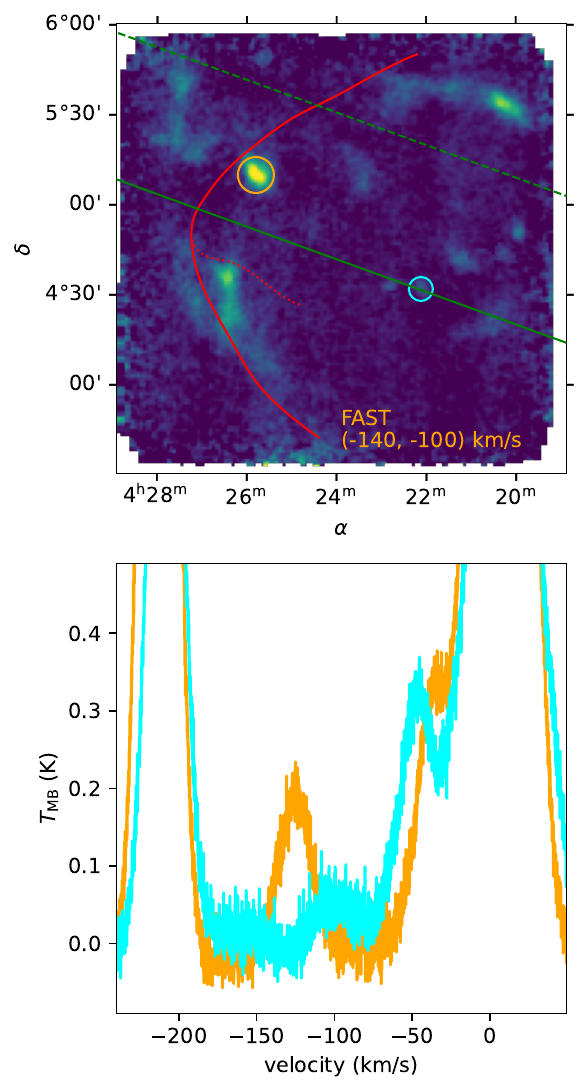}
\caption{
Upper panel: Moment~0 map of the \ion{H}{I} emission for the IVC component, integrated over the velocity range $-140$ to $-100$~$\mathrm{km\,s^{-1}}$. The solid and dashed red curves mark the boundaries of D1, corresponding directly to the blue curves in the upper panel of Fig.~\ref{fig:fronts}. Green lines indicate the two trajectories used to extract the position–velocity slices for Fig.~\ref{fig_pvs} (which are also marked as black lines in Fig.~\ref{fig:HI4PImom0}). Lower panel: Example spectra of the IVC component for two individual clumps (shown in orange and cyan) extracted from the positions marked by the circles of corresponding colors in the upper panel.
}
    \label{fig:ivc}
\end{figure}

\begin{figure*}
    \centering
    \includegraphics[width=0.99\linewidth]{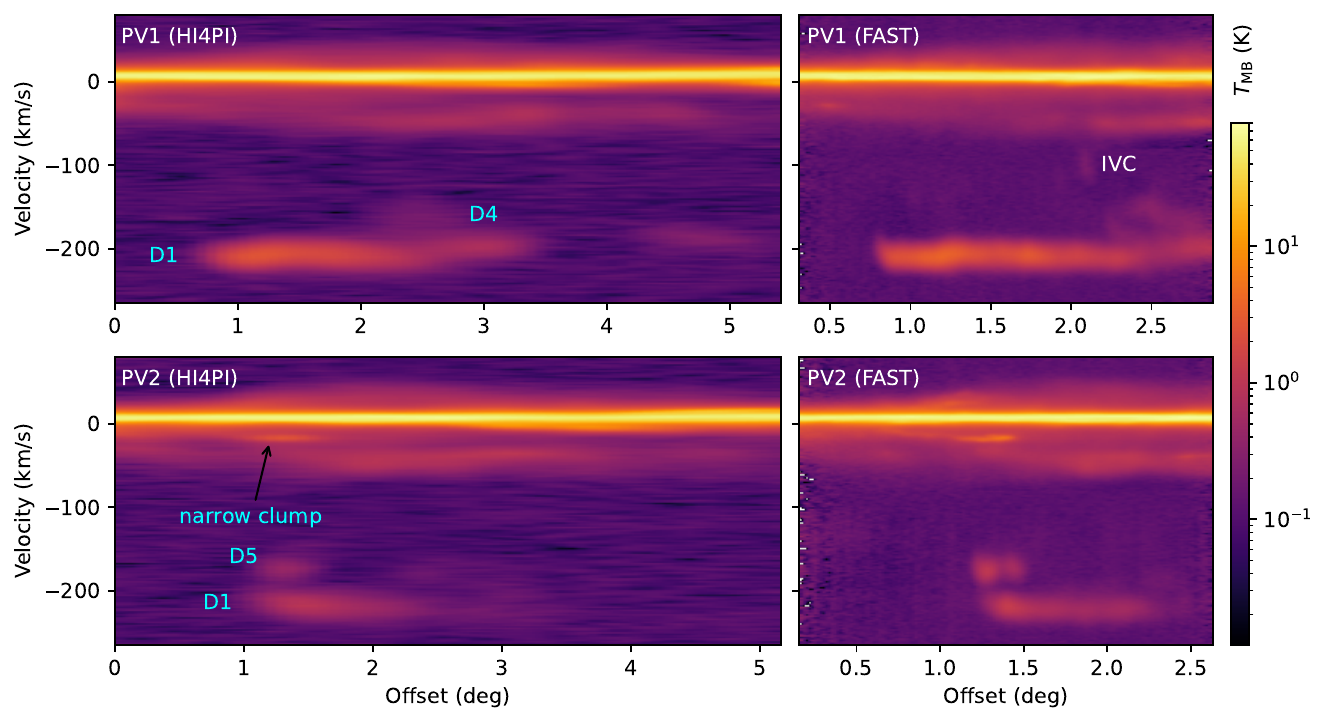}
\caption{Position-velocity (PV) diagrams along two cuts --- PV1 in the upper panels and PV2 in the lower panels --- extracted from the HI4PI (left) and FAST (right) data cubes. The trajectories of PV1 and PV2 correspond to the solid and dashed green lines marked in Fig.~\ref{fig:ivc}, respectively. The respective emission components from the individual droplets (D1, D4, and D5), a narrow foreground clump at $\sim -20\ \mathrm{km\,s^{-1}}$, and the intermediate-velocity cloud (IVC) layer at $\sim -120\ \mathrm{km\,s^{-1}}$ are indicated, with the latter corresponding to the spatial peaks marked by the circles in Fig.~\ref{fig:ivc}.
\label{fig_pvs}
}
\end{figure*}

\subsection{Turbulence perspective} \label{sect_turb}

In the general case where deformation and rotation coexist, 
the Bernoulli balance (Eqs.~\ref{eq_theBc} and \ref{eq_rotBc}) across streamlines for a single eddy can be written in the effective form
\begin{equation}
\alpha |\mathbf{V}|^2 + T \ln \rho + \Phi = \text{const}, \label{eq_alpha}
\end{equation}
where $\alpha$ represents the net effect of local deformation and rotation. 
Deformation-dominated flows correspond to $\alpha \to 1/2$, while rotation-dominated flows correspond to $\alpha \to -1/2$. 
In complex flow fields, $\alpha$ can be interpreted as an integral quantity capturing the combined effect of multiple interacting eddies.

We now interpret turbulence support using this Bernoulli framework. On large scales, near the turbulence driving scale, the flow is quasi-steady and often dominated by coherent rotational or shearing structures. In this regime, the Bernoulli-like balance reads
\begin{equation}
   -\frac{1}{2} |\mathbf{V}_{\rm ls}|^2 + T \ln(\rho) + \Phi = \text{const},
   \label{eq_lsbo}
\end{equation}
where the negative kinetic term provides centrifugal support against local pressure enhancements, sustaining turbulent motions.

At smaller scales, the flow becomes increasingly isotropic and intermittent. Following Kolmogorov scaling \citep{1995tlan.book.....F}, the eddy turnover time $l/v_l$ decreases with scale $l$, meaning small-scale motions cannot maintain long-lived coherent rotation. When averaged over a large-scale background density $\bar{\rho}$, these small-scale velocity fluctuations largely cancel out directionally, leaving a net isotropic kinetic contribution that acts as an effective thermal support:
\begin{equation}
     T_{\rm eff} = T + |\mathbf{V}_{\rm ss}|^2. \label{eq_Teff}
\end{equation}
To reconcile this effective thermodynamic description with our general Bernoulli framework (Eq.~\ref{eq_alpha}) and obtain a macroscopically smoothed balance of the form
\begin{equation}
    T_{\rm eff} \ln(\bar{\rho}) + \Phi \simeq \text{const},
\end{equation}
the net Bernoulli coefficient $\alpha_{\rm net}$ must scale directly with the background density field, such that
\begin{equation}
   \alpha_{\rm net} \sim \ln(\bar{\rho}).
   \label{eq_lnrhoformat}
\end{equation}
This indicates that variations in the background log-density $\ln(\bar{\rho})$ contribute monotonically to the effective energy balance.

At local scales, velocity fluctuations manifest as random, isotropic increments driven by a cascade of nested, un-correlated small-scale eddies. In this regime, the local Bernoulli coefficient $\alpha$ is continually refreshed across logarithmic energy intervals $\ln(T_{\rm eff}/T)$, which physically trace the number of cascading steps down to the dissipation scale. Crucially, because each nested eddy step introduces an independent, random directional re-orientation of the local velocity vector relative to the background streamlines, the cumulative effect behaves as a spatial random walk rather than a monotonic function. By the central limit theorem, while the macroscopic background log-density scales monotonically as in Eq.~\ref{eq_lnrhoformat}, the variance of these stochastic directional increments scales with the square root of the logarithmic energy interval.
That is,
\begin{equation}
    \alpha_{\rm net} = \mathcal{N} \, \sqrt{\ln\!\left(\frac{T_{\rm eff}}{T}\right)},
\end{equation}
where $\mathcal{N}$ is a standard Gaussian random variable. Substituting this expression into Eq.~\ref{eq_alpha} leads to
\begin{equation}
 T \ln(\rho) + \mathcal{N} \, \sqrt{\ln\!\left(\frac{T_{\rm eff}}{T}\right)}\, c_s^2 = \text{const},
\end{equation}
where large-scale variations $\Phi$ are omitted for clarity. This formulation naturally yields a lognormal density distribution with a log-density variance of
\begin{equation}
\sigma_s^2 = \ln(1+\mathcal{M}^2),
\end{equation}
where $\mathcal{M} = |\mathbf{V}_{\rm ss}| / V_T$ is the turbulent Mach number. This result is fully consistent with the established thermodynamic framework of compressible turbulence \citep{2025arXiv250220458L}.

In summary, this subsection unifies our view of dynamic support under a single, cohesive Bernoulli framework. On large scales, ordered flow loops and coherent velocity structures provide macroscopically centrifugal support. As energy cascades to smaller scales, these coherent structures dissolve into isotropic fluctuations, ultimately providing pressure-like support by inducing a stochastic lognormal density field.

\subsection{Dynamic patterns under anisotropic pressure} \label{sec_pattern}
A gas clump subjected to overpressure along a single direction, such as a uniform gravitational or equivalent acceleration field, can maintain a steady state through a standard exponential density profile. When the external force is complex and multi-directional, no static density configuration can fully balance the resulting uneven stresses, forcing internal gas flows to redistribute mass and energy. Although the external interaction is complex, it remains largely static over time. Consequently, organized, directional flow loops are more effective at resisting these steady, uneven forces than isotropic, random turbulence (Sect.~\ref{sect_turb}). These forces naturally spin the internal gas into coherent ring-like or toroidal loops (Fig.~\ref{fig:dynamicpattern}). Within our Bernoulli framework, this spinning motion drives the steady energy balance (Eqs.~\ref{eq_lsbo}), generating the directional centrifugal forces required to counteract the uneven overpressure.

These internal motions are further modulated by the surrounding medium, where time-dependent external flows continuously shape the clump’s boundary. External flows characterized by a high Reynolds number (Sect.~\ref{sec_dropfit}) exert strong, fluctuating ram pressures on the droplet surface. These pressure variations induce small-scale density perturbations across the axisymmetric, parabolic surface of the droplet, carving out alternating ridges and depressions. When viewed in projection, these surface modulations manifest as strip-like density structures aligned roughly with the droplet’s main axis (Fig.~\ref{fig:dynamicpattern}). Within the broader Bernoulli context detailed in Sect.~\ref{sec_droplets}, these localized structures reflect the conversion of fluctuating external kinetic energy into steady, high-density boundary enhancements.

Interactions between neighbouring droplets can further complicate both internal and external flows. Admittedly, 2D projection effects mean we cannot completely rule out the effect of inhomogeneities in the surrounding medium. However, because these neighboring droplets share very similar velocities and heading directions, a direct physical interaction remains a preferred explanation (Sect.~\ref{sec_droplets}). In general, such interactions may simply result from intersecting trajectories. In the case of droplet D1 in AC-III, however, it is influenced by several neighbouring droplets located in different directions. This geometry suggests the existence of a specific mechanism that can effectively push neighbouring droplets together. One possibility is the Bernoulli effect: when the flow accelerates between adjacent droplets, a low-pressure region is created that draws the droplets closer (Eq.~\ref{eq_theBc}; Fig.~\ref{fig:dynamicpattern}). According to the Bernoulli formulation (Eq.~\ref{eq_theBc}), these interactions can trigger internal loops, enhancing the complexity of gas motions and contributing to the overall dynamic behavior of the system.

\section{Discussion} \label{sec_discuss}

\subsection{AC-III dropping through the WIM}\label{dis_halo}
The droplet shape of AC-III indicates that it is actively sculpted by the ram pressure of external flows with an ambient density of $n_{\rm ext}\sim 10^{-3}\ \mathrm{cm^{-3}}$ (Sect.~\ref{sec_aciiifit}). This value is one to two orders of magnitude higher than the typical density of the hot halo, which hovers around $10^{-5}$ to $10^{-4}\ \mathrm{cm^{-3}}$ \citep{2012ApJ...756L...8G,2015ApJ...800...14M} at Galactocentric radii exceeding $10$~kpc where standard halo HVCs reside. The general rarity of these droplet-like features across the sky strongly suggests that the ram pressure of this diffuse, spherical hot halo is simply too weak to shape infalling clouds. Instead, our derived value for $n_{\rm ext}$ is remarkably consistent with the denser, disk-stratified warm ionized medium (WIM; \citealt{1991ApJ...372L..17R,2008PASA...25..184G}), which is structurally bounded by the Galactic disk gravity and magnetic fields \citep{1990ApJ...365..544B}. 
Bounded by a plane-parallel geometry, the WIM density drops off exponentially as a function of vertical height $z$ above the disk. Adopting a standard midplane WIM density of \(n_0 \sim 0.01\ \mathrm{cm^{-3}}\) and a vertical scale height of \(H_{\rm WIM} \sim 1\ \mathrm{kpc}\) \citep{2008PASA...25..184G}, the WIM density at the specific vertical altitude of AC-III ($z = d \sin b$, taking the distance $d = 5$~kpc) is calculated as:
\begin{equation}
    n_{\rm WIM} = n_0 \exp\left(-\frac{d \sin b}{H_{\rm WIM}}\right) \sim 8\times 10^{-4}\ {\rm cm^{-3}}, 
\end{equation}
which tightly matches our independent estimate in Eq.~\ref{eq_esnext}. Therefore, AC-III is likely experiencing ongoing hydrodynamic deceleration due to external drag from this denser WIM substrate. Because its observed infall velocity, \(\|\mathbf{v}_{\rm ram}\|\), is comparable to the local free-fall velocity dictated by Galactic disk and halo gravity, the cloud must be situated several kiloparsecs above the plane, catching the exact physical boundary where it transitions from the spherical hot halo into the disk-stratified Galactic WIM layer.

This physical scenario carries several immediate implications for the broader study of halo gas evolution. It implies that most classical HVC complexes are located outside the Galactic WIM layer (Fig.~\ref{fig_sketch-large}), which explains why droplet-like morphologies are so rare in current wide-field surveys. It also suggests that infalling HVCs remain dominated by a warm neutral medium (WNM) phase until they actually penetrate and are compressed by this denser WIM environment. However, validating these claims requires a more comprehensive investigation of ram-pressure-influenced HVCs. Substructures embedded within larger HVC complexes occasionally exhibit stripped morphologies or hints of head--tail structures \citep[e.g.,][]{2021AAS...23752709B}; higher-resolution observations are necessary to clearly distinguish between isolated droplets and these nested environments. Furthermore, many compact HVCs have only been surveyed under coarser angular resolution or with limited sensitivity \citep[e.g.,][]{2002A&A...391..713K,2005A&A...432..937W}. Because most compact HVCs possess smaller angular sizes than AC-III, it remains uncertain whether, or how many of them, would resolve into droplet-like morphologies with interferometers \citep[e.g.,][]{2014A&A...563A..99F}. Additionally, for the strip-like patterns detected by interferometers in HVCs \citep[e.g.,][]{2017ApJ...834..126B}, while tidal forces may play an important role in shaping their global morphology \citep[e.g.,][]{2005A&A...436..101W}, ram pressure likely also contributes significantly to their localized stripping. High-resolution imaging of known droplets like AC-III would provide an essential baseline to verify their differences from more distant HVCs or post-interaction clouds like the Smith Cloud \citep[e.g.,][]{2008ApJ...679L..21L,2009ApJ...707.1642N}, thereby confirming whether unique signatures of strong ram-pressure interactions associated with the WIM boundary can be resolved via interferometry. In summary, these claims can only be systematically tested as dedicated high-resolution mapping data become available. More importantly, systematic mapping of these droplet head shapes could allow us to use HVCs as direct indicators to place competitive observational constraints on the scale height and density profile of the Galactic WIM itself.

The validity of these broader implications also inherently depends on the robustness of our physical interpretation. While our simplified analytic framework does not account for every physical mechanism operating at the halo--disk interface, it captures the primary hydrodynamical behavior of the system. In a realistic system, a variety of additional physical processes introduce inherent uncertainties; for instance, a primordial helium fraction, an enveloping ionized sheath \citep{2004ApJ...602..738F}, or a stabilizing dark matter halo \citep{2009ApJ...707.1642N} would alter the cloud's effective mass and structural lifetime. Similarly, ambient magnetic fields \citep{2002A&A...391..713K} and localized overdensity clumps within the hot halo may modulate the ram-pressure balance \citep{2009ApJ...698.1485H}. Beyond these ambient factors, the HVC itself likely possesses a complex multi-phase architecture \citep[e.g.,][]{2006A&A...455..481K}, opening up the possibility that its diffuse outer ionized phase is being actively shocked and heated through distant interactions with the neighboring IVC layers (see further discussion in Sect.~\ref{sec_IVC}). Nevertheless, a dynamic interaction with the stratified WIM remains the most direct and compelling physical candidate to explain the droplet-like morphology of the cloud.

\begin{figure}
    \centering
    \includegraphics[width=0.99\linewidth]{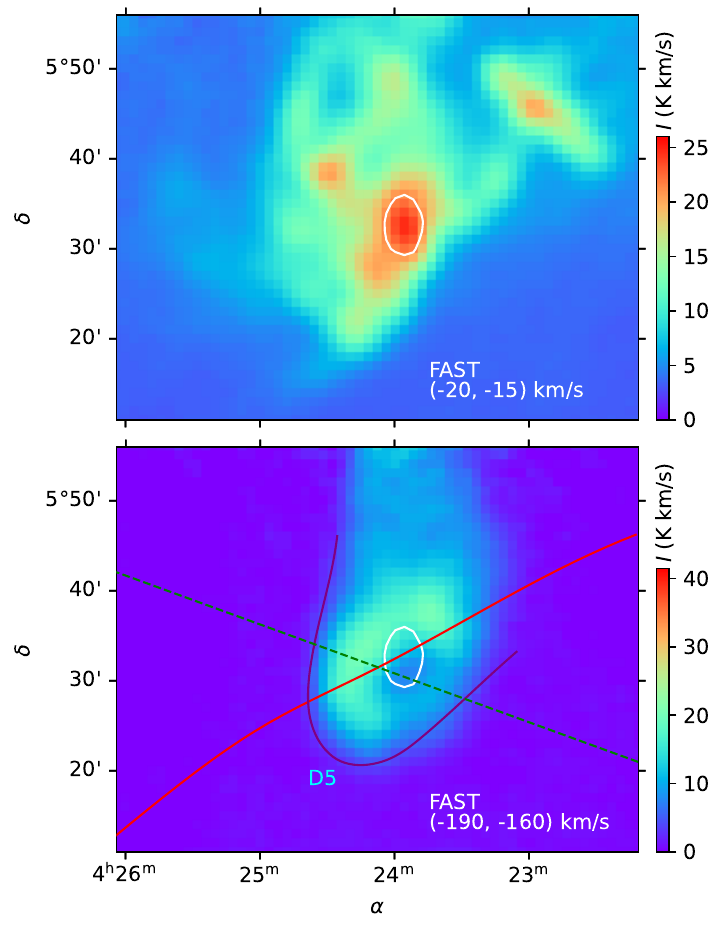}
\caption{
Upper panel: Moment~0 map of the foreground narrow clump, which is spatially overlapping with the projected boundary of D5 on the sky (see Fig.~\ref{fig_pvs}), with the white contour set at 20~$\mathrm{K\,km\,s^{-1}}$. Lower panel: Moment~0 map of droplet D5, overlaid with a white contour identical to that displayed in the upper panel to facilitate direct structural comparison. The red and purple curves mark the droplet heads of D1 and D5, respectively (see also the upper panel of Fig.~\ref{fig:fronts}). The green dashed line indicates the trajectory of the position–velocity cut PV2 (see Fig.~\ref{fig:HI4PImom0}).
}
    \label{fig:narrowcomp}
\end{figure}

\subsection{Reynolds number of external flows} \label{sec_external}

The ambient gas, whether residing within the warm ionized medium (WIM) or the extended hot halo (Fig.~\ref{fig_sketch-large}), consists of a fully ionized hydrogen plasma whose kinematic properties are governed by Coulomb collisions that determine its dynamic viscosity. The dynamic viscosity ($\eta$) of such a plasma can be expressed as \citep{1953PhRv...89..977S}:
\begin{equation}
\eta = 2.21 \times 10^{-15} \, \frac{T^{5/2}}{Z^4 \, \ln \Lambda} \quad 
\left[ \mathrm{g \, cm^{-1} \, s^{-1}} \right],
\end{equation}
where $T$ is the gas temperature in Kelvin, $\ln \Lambda$ is the Coulomb logarithm (typically $\sim 30$), and $Z$ is the particle charge number ($Z = 1$ for a pure hydrogen plasma). 

The Reynolds number ($\mathrm{Re}$) characterizing this external flow is defined as:
\begin{equation}
\mathrm{Re} = \frac{V L \rho}{\eta} = \frac{V L n_{\rm ext} m_p}{\eta},
\label{eq_Rey}
\end{equation}
where $L \sim H$ represents the characteristic linear scale of a droplet and $m_p$ is the proton mass, which dominates momentum transport relative to electrons. Assuming a distance scale factor $d_5 = d / 5\ \rm kpc$, Eq.~\ref{eq_Rey} can be scaled as:
\begin{equation}
\mathrm{Re} \sim 10 \, \left[ H \, d_5 \right] \, \left[ \frac{T}{10^6\ \rm K} \right]^{-5/2}
\left[ \frac{V_{\rm ext}}{200\ {\rm km \, s^{-1}}} \right]
\left[ \frac{n_{\rm ext}}{10^{-4}\ {\rm cm^{-3}}} \right],
\end{equation}
where $H$ is the angular cross-section of the droplet in degrees. Because $\eta \propto T^{5/2}$, the Reynolds number decreases dramatically as the gas temperature increases. 

For a typical droplet with an angular scale of $H \sim 1^\circ$ traveling at $V_{\rm ext} = 200\ \mathrm{km\,s^{-1}}$, adopting parameters characteristic of the WIM ($n_{\rm ext} = 10^{-3}\ \mathrm{cm^{-3}}$, $T \sim 10^4\ \mathrm{K}$) yields a highly turbulent regime with $\mathrm{Re} \sim 10^7$. Conversely, applying parameters typical of the diffuse hot halo ($n_{\rm ext} = 10^{-4}\ \mathrm{cm^{-3}}$, $T \sim 10^6\ \mathrm{K}$) yields a viscous regime with $\mathrm{Re} \sim 10$. 

Furthermore, the thermal velocity dispersion at $10^6\ \mathrm{K}$ is $\sim 100\ \mathrm{km\,s^{-1}}$, which is comparable to the typical bulk velocity of an HVC, whereas the thermal dispersion at $10^4\ \mathrm{K}$ is only $\sim 10\ \mathrm{km\,s^{-1}}$, which is substantially lower than the bulk infall motion. Consequently, a droplet tends to move smoothly through the hot halo under transonic or subsonic conditions where the fluid exhibits a moderate Reynolds number. However, upon plunging into the colder, denser WIM regime, the cloud transitions to a highly supersonic state where it experiences strongly turbulent flows. The net deceleration imparted by ram pressure depends heavily on how the fluid streamlines detach from the cloud surface. In the high-Reynolds-number WIM regime, the effective cross-sectional interaction area $A$ (Eq.~\ref{eqFram}) extends directly across the droplet surface, driving both the efficient deceleration and the prominent surface fluctuations observed throughout the AC-III complex. Note that this localized hydrodynamical treatment omits ambient magnetic fields within the WIM, which are expected to suppress and damp these surface shear instabilities \citep{2002A&A...391..713K}.

\subsection{Foreground and IVC components} \label{sec_IVC}

In Sects. \ref{dis_halo} and \ref{sec_external}, we discussed the probability of an active dynamical interaction between AC-III and the warm ionized medium (WIM). However, the overall line-of-sight environment is kinematically rich, containing distinct intermediate-velocity cloud (IVC) layers and cold foreground neutral components alongside the primary HVC (Sects. \ref{sec_results} and \ref{sec_foretemp}). To build a complete physical picture of the system, it is necessary to check for possible structural or kinematic connections between these lower-velocity components and the fast-moving cloud complex. Because AC-III appears to be transiting the WIM at a relatively low altitude above the Galactic disk, the physical likelihood of dynamic coupling across these distinct velocity phases is inherently elevated. By performing a comparative morphological analysis in the spatial domain and searching for velocity bridge features in the spectral domain \citep[e.g.,][]{1996A&A...308L..37P}, we can systematically evaluate the relationship among these distinct \ion{H}{I} phases.

Our spatial morphological comparison begins with Fig.~\ref{fig:ivc}, which displays the moment~0 map of the IVC component integrated over the velocity range of $-140$ to $-100\ \mathrm{km\,s^{-1}}$. Spatially, the IVC emission traces a prominent, shell-like boundary layer that mimics the curvature of, and partially envelops, the droplet head of D1. High-resolution FAST data reveal that this morphologically aligned shell is fragmented into compact, discrete clumps with angular sizes comparable to the telescope beam, several of which lie precisely along the middle axis in the projected plane of D1 (green solid line in Fig.~\ref{fig:ivc}, PV1). These small-scale sub-structures are completely unresolved in the lower-resolution HI4PI survey, underscoring the critical role of high-sensitivity, high-angular-resolution spatial mapping in identifying such boundary interfaces. To evaluate the kinematic relationship, our spectral search focuses on the position-velocity (PV) diagram extracted along the PV1 trajectory (upper panels of Fig.~\ref{fig_pvs}), alongside example spectra of the IVC clumps (lower panel of Fig.~\ref{fig:ivc}). Spectrally, these clumps occupy a velocity regime closer to the stable foreground Galactic disk than to the bulk HVC motion. Although the spatial alignment between the IVC shell and the head of D1 is striking, the spectral search reveals a prominent $\sim 50\ \mathrm{km\,s^{-1}}$ velocity gap between the two components, confirming the absence of a continuous, direct kinematic bridge. Viewed within the context of interstellar drag, this configuration points to a spatially aligned but indirectly mediated interaction: the infalling HVC may be driving a supersonic shock wave into the invisible WIM precursor, which subsequently acts as a physical piston that dynamically compresses, radiatively cools, and condenses the ambient gas into the observed neutral IVC clumps.

A similar spatial and spectral configuration is observed near the D5 component. In the spatial domain, a compact clump centered at $\sim -20\ \mathrm{km\,s^{-1}}$ (hereafter the ``narrow clump''; see the lower panel of Fig.~\ref{fig_pvs}) outlines a contour that closely mirrors the morphology of D5, including a convex front edge facing west (Fig.~\ref{fig:narrowcomp}). Crucially, the peak intensity of this narrow foreground clump coincides perfectly with H4, the prominent emission hole within the main body of D5 (Fig.~\ref{fig:fronts}). The narrow line width of $\sim 3\ \mathrm{km\,s^{-1}}$ indicates that this foreground feature is dominated by the cold neutral medium \citep[CNM,][]{2003ApJ...586.1067H}. Given that no corresponding structures are identified in the broader major or minor foreground \ion{H}{I} components—which span standard velocity ranges of $-10$ to $20\ \mathrm{km\,s^{-1}}$ and $-50$ to $-20\ \mathrm{km\,s^{-1}}$, respectively—this highly localized spatial alignment is unique. In the spectral domain, however, the PV diagram reveals an emission gap that confirms the absence of any connecting bridge feature, indicating no direct collision or mass transfer between them. Nevertheless, the precise morphological matching and the spatial interlocking of the CNM peak within the H4 cavity are difficult to attribute to mere chance. Instead of dismissing this alignment as a simple projection effect, this structure could trace a multi-phase hydrodynamical process wherein the ram-pressure shadow or the localized wake trailing behind D5 alters the local pressure balance of the WIM. Such a localized pressure drop could trigger thermal instabilities in the ambient gas, causing it to quickly condense into the narrow CNM clump observed directly along the line of sight.

These spatial and spectral investigations of the droplet shapes, internal loops, and broken shells throughout AC-III help us understand how high-velocity clouds change as they enter the inner halo. As an infalling cloud plows through the WNM, its front edge is heavily squeezed by the surrounding gas, a process that can easily trigger turbulent instabilities that strip gas away from the core and shred it into intertwined filaments. However, because we do not see continuous velocity bridges in our data, these disruptions must still be in an early stage, or the clouds are not in direct contact. Instead, the interaction is indirect and mediated entirely by the surrounding medium, possibly stabilized by magnetic field lines that act as a shield to keep the droplet head smooth. To fully understand how energy moves between these layers and clear up line-of-sight perspective confusion, more data are required, including precise distance measurements, high-sensitivity H$\alpha$ maps to image the invisible WIM layer, and ultraviolet absorption data against background stars to map the temperature and ionization of the gas.

\section{Summary and conclusions}\label{sec_summary}
The high-resolution FAST \ion{H}{I} observations reveal unprecedented internal structures within HVC AC-III. We have systematically analyzed the morphologies, multi-phase structures, and internal dynamical patterns of the complex. Our primary findings are summarized as follows:

\begin{enumerate}
    \item The HVC AC-III is structurally composed of multiple coherent subclumps (denoted D1 through D6), each characterized by a nearly constant Gaussian line width of $\sim 20~\mathrm{km~s^{-1}}$ and a mean volume density of $\sim 0.2~\mathrm{cm^{-3}}$. In contrast, the global velocity distribution of the complex is broad, spanning from $-220$ to $-180~\mathrm{km~s^{-1}}$. This kinematic configuration implies that each subclump is a coherent parcel of subsonic or transonic warm neutral medium (WNM), with the bulk of the system's kinetic energy stored in clump-to-clump relative motions.
    
    \item These subclumps exhibit symmetric parabolic morphologies, indicating that they are droplets confined by external ram pressure. Under a steady-state analytical framework, the deceleration induced by external drag is mathematically equivalent to a linear potential that compresses the neutral gas toward the forward droplet head, matching our observed profiles.

    \item A closed-form steady-state model fit yields a peak droplet-head density of $\sim 2~\mathrm{cm^{-3}}$, which is roughly an order of magnitude higher than the average volume density of the cloud. This high density requires compression by an external ambient medium with a density of $\sim 10^{-3}~\mathrm{cm^{-3}}$, which is consistent with that of the disk-stratified WIM at the estimated vertical height ($z$) of AC-III, assuming a distance of $5~\mathrm{kpc}$. Although the inclination angle ($\theta$) cannot be uniquely constrained, a significant tilt from the line of sight is necessary, implying a substantial perpendicular velocity component.

    \item The individual droplets within AC-III are likely shaped by ram pressure as they enter the Galactic WIM layer. Because such droplet-like morphologies are generally rare in wide-field surveys, the majority of classical HVCs are likely located outside this layer. From a dynamical perspective, we suggest that HVCs generally reside at moderate Galactocentric distances ranging from several to approximately ten kiloparsecs, placing them within the inner halo rather than in the distant outer halo or intergalactic space.

    \item The primary droplet, D1, exhibits localized departures from axial symmetry. This distortion is likely driven by hydrodynamical interactions with neighboring subclumps. Furthermore, detailed internal structures including linear ridges, rings, and cavities are resolved within the droplet bodies, suggesting the presence of complex internal gas dynamics responsible for these structural features.

    \item An intermediate-velocity cloud (IVC) component, spanning velocities from $-150$ to $-100~\mathrm{km~s^{-1}}$, traces a distinct shell-like morphology that partially envelops the droplet head of AC-III. While position-velocity diagrams confirm the absence of a direct spectral bridge between the IVC and HVC phases, indirect physical coupling mediated by large-scale perturbations, compression zones, or density waves within the surrounding WIM medium cannot be excluded.

    \item Within the framework of the Bernoulli equation, we present a comprehensive physical model of ram-pressure–induced droplets. Under uniform ram pressure, a droplet naturally develops a parabolic boundary layer and an exponential density profile, while external drag tends to drive neighboring droplets to merge. Due to the high Reynolds number of the interaction, external turbulence triggers density fluctuations along the boundary layer and the middle symmetry axes, manifesting as the observed linear ridges. Finally, internal fluid loops arise in response to anisotropic external pressures and close-range droplet interactions, forming stable, long-lived ring-like configurations. These fluid loops can persist over prolonged timescales and drive secondary internal turbulence.
\end{enumerate}

Overall, we suggest that HVC AC-III is actively entering the disk-stratified Galactic WIM layer and being sculpted by external ram pressure into a droplet-like morphology. This complex provides an ideal laboratory for investigating structure formation, turbulence generation, and the multi-phase evolution of halo gas infalling toward the Galactic plane, while serving as a sensitive probe of the physical properties of the ambient circumgalactic medium.

\begin{acknowledgement}
We thank the FAST staff for their assistance during the observations.
X.L. acknowledges support from the Strategic Priority Research Program of the Chinese Academy of Sciences under Grant No. XDB0800303.
We thank the anonymous referee for their helpful comments and suggestions, which have strengthened this work.
\end{acknowledgement}

\bibliographystyle{aa}
\bibliography{AC-III}

\begin{appendix}
\section{Issues in data reduction}\label{sec_rfi}

Figure~\ref{fig:HI4PImom0} shows the moment-0 map of AC-III as revealed by HI4PI
\citep{2016A&A...594A.116H}. 
The four OTF fields observed with FAST are also indicated. 
The FAST observations are described in Sect.~\ref{sec_fastobs}, 
and the data reduction procedure follows \citealt{2025NatAs...9.1366L}.

\begin{figure}[!h]
        \centering
        \includegraphics[width=0.99\linewidth]{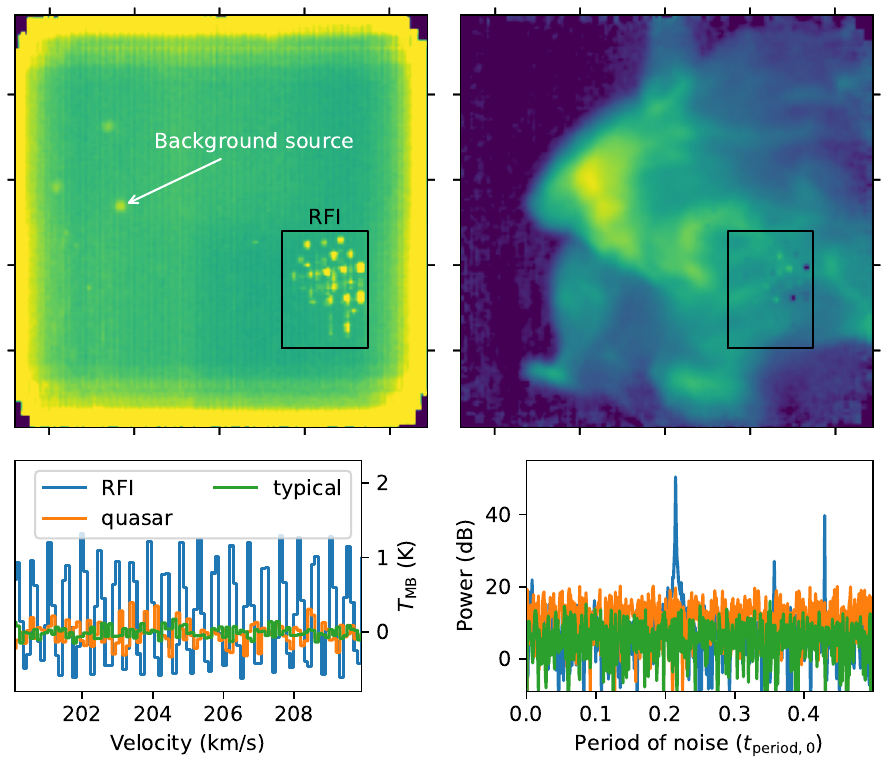}
        \caption{
Upper left: One-$\sigma$ noise map of the RFI-contaminated FAST cube. The black box marks the RFI-affected region, and the white arrow points to the strongest background radio source. 
Upper right: \ion{H}{I} map of AC-III integrated from the same cube. 
Lower left: Example spectra of RFI-affected pixels, pixels near the quasar, and normal pixels. 
Lower right: Fourier transform of the example spectra over 100--200~km~s$^{-1}$ (emission-line-free range), showing noise power versus period. Line colors correspond to the pixels shown in the lower-left panel.
        \label{noiserfi}}
    \end{figure}

The noise map of the FAST data cube, obtained by calculating the standard deviation over the line-free channels, is relatively uniform in the central part of the mapped region, with higher values toward the margins due to limited scanning time and at the positions of bright background radio sources (usually quasars), as expected. In addition, some unexpected high-noise spots are present, which are enclosed by black squares in the upper left panel of Fig.~\ref{noiserfi}. These features are likely caused by radio frequency interference (RFI), which severely affects the \ion{H}{I} integrated intensity map at the corresponding pixels (see upper right panel of Fig. \ref{noiserfi}). Therefore, it is necessary to remove them before further analysis.  

 \subsection{Remove RFI} \label{sect_removrfi}
    
    \begin{figure}
        \centering
        \includegraphics[width=0.95\linewidth]{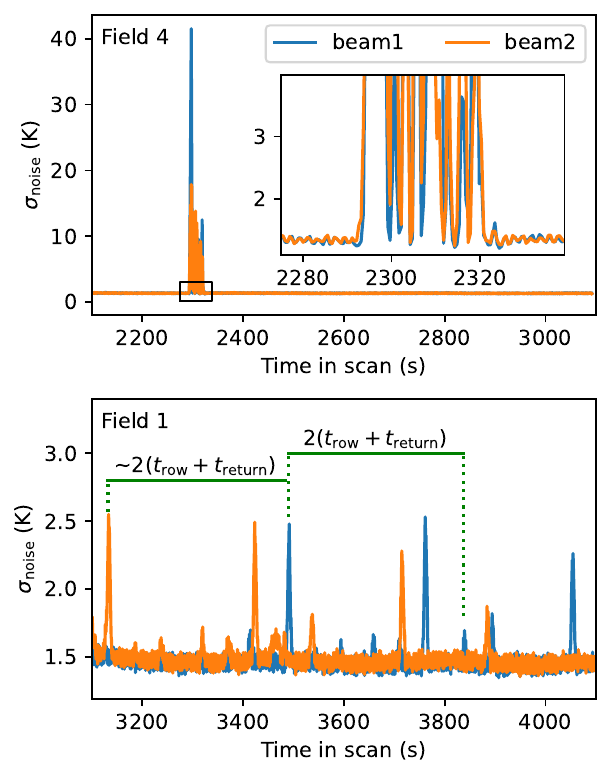}
        \caption{Upper: Noise level versus time for Field~4, with intervals affected by strong RFI highlighted. 
The inset shows a zoomed-in view of the region indicated by the black box. 
Different colors represent different beams. 
Lower: Noise level for a time segment from Field~1, showing the effect of background radio sources. 
Green lines mark noise enhancements separated by time intervals of $2(t_{\rm row}+t_{\rm treturn})$, 
corresponding to the same sources being observed twice by the same beam and by different beams along the same scan direction. 
For successive rows scanned in the opposite direction, the interval between noise peaks varies with the source position along the scan.
        \label{noisetime}}
    \end{figure}

    \begin{figure*}
        \centering
        \includegraphics[width=0.99\linewidth]{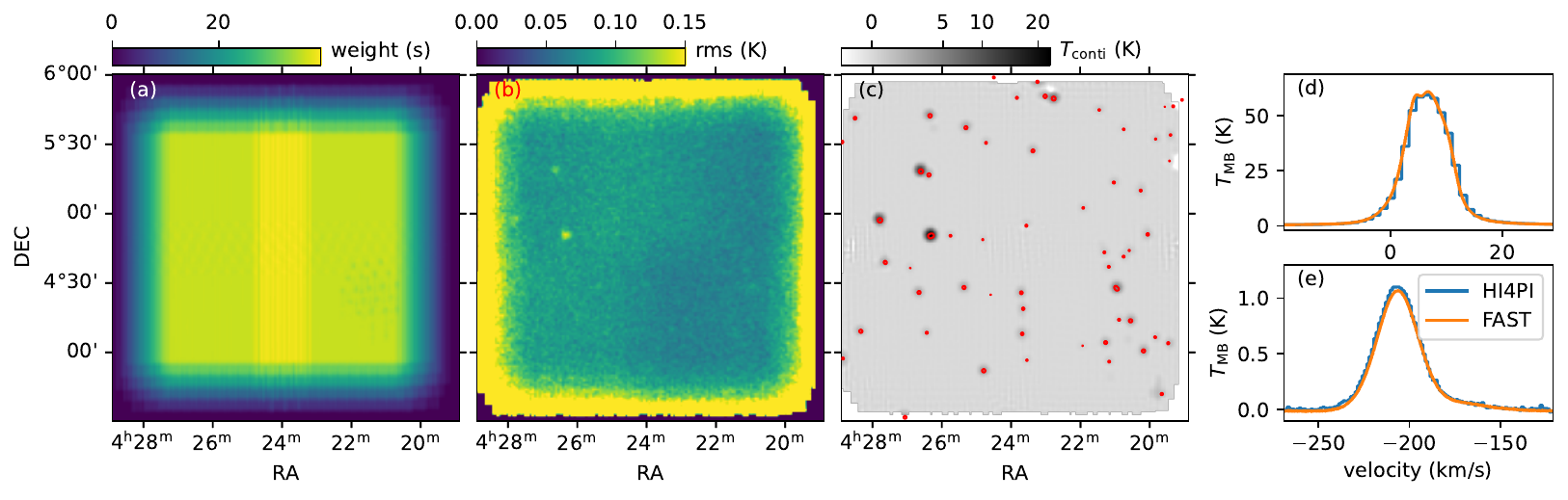}
        \caption{
(a) Weight map of the OTF data after gridding, showing spatial coverage and relative sensitivity across the field.  
(b) One-$\sigma$ noise map of the cube after RFI removal, illustrating noise variations across different regions.  
(c) FAST $L$-band continuum map overlaid with NVSS L-band contours at 15~mJy~beam$^{-1}$, highlighting strong radio sources.  
(d)--(e) Comparison of mean spectra averaged over the observed region from FAST and HI4PI, for the foreground and high-velocity cloud (HVC) components, respectively.
\label{fig:cleancubestat}
        }
    \end{figure*}

The RFI-affected pixels are distributed in a pattern corresponding to the 19 beams of the receiver, suggesting that the interference occurred within a short time interval and simultaneously affected all beams. In the lower left panel of Fig.~\ref{noiserfi}, we compare representative spectra in a line-free velocity window from a pixel affected by RFI, a pixel near a quasar, and a normal pixel. The noise in the RFI-uncontaminated pixels (those near a quasar and normal pixels) appears visually white-noise–like, with the noise near the quasar showing a higher standard deviation. In contrast, the noise at the RFI-contaminated pixel is overwhelmed by a periodic comb function in the frequency domain, with a period of about four to five $\delta f$. The Fourier transform of the noise spectra (within an emission-line-free velocity range of 100–200~km~s$^{-1}$), shown in the lower right panel of Fig.~\ref{noiserfi}, confirms that the RFI-uncontaminated noise is indeed white noise in the period domain, whereas the RFI-contaminated noise has its power concentrated around three periods, with the most significant peak at $t_{\rm period,\ RFI}\sim 0.215\,t_{\rm period,0}\sim 0.45$~ms.
Overall, the noise at certain pixels is strong, intermittent, and exhibits a comb-like pattern in the frequency domain, corresponding to a characteristic period of $\sim 0.45$~ms in the time domain, in contrast to the visually white-noise–like behavior of normal pixels.

We plot the noise level of each sampled spectrum (with a sampling interval of $\sim 0.5$~ms; see Sect.~\ref{sec_fastobs}) as a function of sampling time and visually inspect the samples exhibiting abnormal noise. Only a single short time interval is found to contain obviously abnormal values, as shown in Fig.~\ref{noisetime}. This interval (RFI time interval) lasts about half a minute during the OTF scan of the fourth field. 
The noise contributed by the system temperature is $\sim 1.5$~K, while the noise level can reach up to 40~K during the RFI interval. Note that the noise level also increases when the beams scan around background radio sources. The strongest radio source contributes to a noise level increase of $\sim 1$~K. 
The noise increment caused by background radio sources can be easily recognized based on several characteristics: 
(1) the increase is not as strong as that induced by RFI; 
(2) the affected time interval is narrow, corresponding to the beam size divided by the scanning speed; 
(3) when the row separation is small, a radio source is covered by several successive rows of each beam, with every two visits separated by $2(t_{\rm row}+t_{\rm return})$; 
(4) similar increments occur for all beams, but with a time delay, e.g., the increment in beam 2 is delayed by about $2(t_{\rm row}+t_{\rm return})$ compared to beam 1 (see lower panel of Fig.~\ref{noisetime}).
Here, $t_{\rm row}$ is the scanning time of each row (120 s).

In this work, to obtain the RFI-free merged cube, we simply discard the data within the RFI-contaminated time interval. Note that, since the 19 beams visit the same location at different times, a location affected by RFI in one beam can still be covered by the other 18 beams during non-RFI intervals. Therefore, the impact on the effective integration time and sensitivity is minimal.

\subsection{Verification of RFI-cleaned data cube} \label{sect_checkcube}

The weight distribution from the OTF gridding is uniform across the mapped field (Fig.~\ref{fig:cleancubestat}a), indicating that the removal of RFI-contaminated data does not affect the coverage. The rms noise of the final cube within the central $1.5\degr \times 1.5\degr$ region is $75 \pm 8$~mK at the native velocity resolution of $\sim 0.1$~km~s$^{-1}$ (Fig.~\ref{fig:cleancubestat}b), and the RFI-induced spots on the noise map are removed as expected. The continuum map extracted from the dataset shows good agreement with the NVSS L-band continuum in both morphology and integrated flux density (Fig.~\ref{fig:cleancubestat}c), with discrepancies within 5\%. A two-dimensional Gaussian fit to the brightest background continuum source (upper left panel of Fig.~\ref{noiserfi}) yields a FWHM beam size of 3.2\arcmin\ and a peak brightness temperature of 16.36~K, corresponding to a flux density of 1.03~Jy~beam$^{-1}$, consistent with the NVSS measurement of 1.08~Jy within 5\% \citep{1998AJ....115.1693C}. The mean H\,{\sc i} spectrum averaged over the observed field is likewise in good agreement with the HI4PI survey \citep{2016A&A...594A.116H} in both spectral profile and absolute intensity (panels d--e of Fig.~\ref{fig:cleancubestat}).

\section{Estimation of \ion{H}{I} temperature}\label{sec_foretemp}
\begin{figure}
    \centering
    \includegraphics[width=0.95\linewidth]{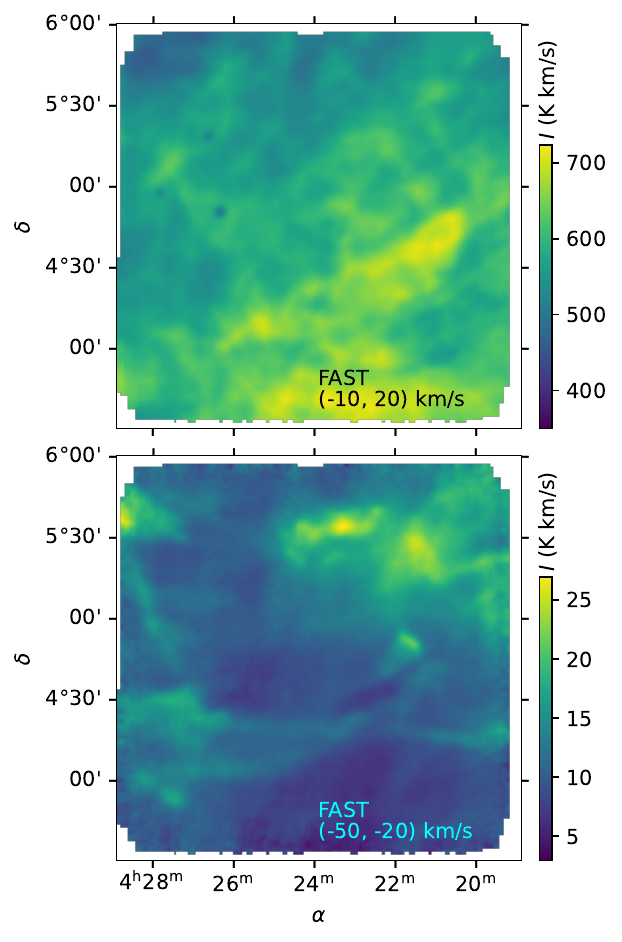}
    \caption{ Foreground \ion{H}{I} emission in the region of AC-III, integrated over the velocity ranges indicated in the lower right corner of each panel.  }
    \label{fig:forg}
\end{figure}

Fig.~\ref{fig:forg} shows the foreground \ion{H}{I} emission of the AC-III region.
On the foreground ($-10$ to 20~km\,s$^{-1}$) \ion{H}{I} moment~0 map (upper panel of Fig.~\ref{fig:forg}),
absoprtion dips against strong background continuum sources (Fig.~\ref{fig:cleancubestat}) are clearly seen.
To quantitatively describe the absroption feature, we apply a two-dimensional Gaussian profile,
\begin{equation}
    f = A \left( 1 - A_r \exp\left(-\frac{r^2}{2\sigma^2}\right) \right),
\end{equation}
to the foreground ($-10$ to 20~km\,s$^{-1}$) \ion{H}{I} moment~0 map around the brightest background continuum source (indicated by the white arrow in Fig.~\ref{noiserfi}). Here, $r$ represents the angular distance from the absorption peak and $\sigma$ denotes the characteristic angular width of the feature. This fit yields an absorption dip depth ratio of $A_r = 0.18$. At the FAST angular resolution, the peak brightness temperature of this brightest background continuum source is measured as $T_c = 16.36$~K (Sect.~\ref{sect_checkcube}).

The observed line brightness temperature ($T_r$) is governed by the standard radiative transfer equation \citep[e.g.,][]{2011piim.book.....D}:
\begin{equation}
    T_r = (T_s - T_c)(1-e^{-\tau}), \label{eq_rtrans}
\end{equation}
where $T_c$ is the brightness temperature of the background continuum, while $T_s$ and $\tau$ are the excitation (spin) temperature and the optical depth of the foreground \ion{H}{I} emitter, respectively. Here, the contribution from any foreground continuum emission is neglected. Eq.~\ref{eq_rtrans} holds valid under the assumption that both $T_r$ and $T_c$ are measured within the same beam size, and that $\tau$ remains approximately uniform across the beam area. Outside the absorption dip where the background continuum falls to zero ($T_c = 0$), the line brightness temperature simplifies to
\begin{equation}
    T_r' = T_s (1-e^{-\tau}). \label{eq_rtranssimple}
\end{equation}
The relative depth of the absorption dip can then be expressed from Eqs.~\ref{eq_rtrans} and \ref{eq_rtranssimple} as
\begin{equation}
    A_r = \frac{T_r' - T_r}{T_r'} = \frac{T_c}{T_s}. \label{eq_Tss}
\end{equation}

By inserting our measured values into Eq.~\ref{eq_Tss}, the spin temperature of the foreground \ion{H}{I} is estimated to be
\begin{equation}
    T_{\rm s,\,f} \sim \frac{T_c}{A_r} \sim 90~\mathrm{K}.
\end{equation}
This value is fully consistent with the characteristic temperatures of the cold neutral medium (CNM; \citealt{2003ApJ...586.1067H,2009ARA&A..47...27K}). Given that the peak line brightness temperature of the foreground \ion{H}{I} is $T_r \sim 60$~K, its optical depth can be evaluated using
\begin{equation}
    \tau = -\ln\left(\frac{T_s-T_r}{T_s}\right),
\end{equation}
which yields $\tau \sim 1$. This implies that the foreground CNM can remain considerably optically thick even at high Galactic latitudes and toward the anti-Galactic center. This result aligns with global interstellar medium frameworks showing that high-latitude \ion{H}{I} emission is frequently saturated, with optical depths often reaching or exceeding unity \citep{2015ApJ...798....6F}. Physically, this unisolated, dense foreground material is likely associated with the cold, swept-up atomic gas residing on the shell walls of the expanding Local Bubble surrounding the solar neighborhood \citep{2022Natur.601..334Z}. Fortunately, the HVC \ion{H}{I} emission will not be absorbed by the foreground 
\ion{H}{I} due to their large velocity separation.

In contrast, no absorption features against the background continuum sources are detected for the high-velocity cloud (HVC) AC-III. On the moment~0 map of HVC AC-III, the signal-to-noise ratio at the position of the brightest continuum source is 135. This yields a $3\sigma$ confidence upper limit on the dip ratio of $A_r < 1/45$. Consequently, the lower limit on the \ion{H}{I} spin temperature of the HVC component is derived as
\begin{equation}
    T_{\rm s,\,HVC} > 45 \times T_c \sim 700~\mathrm{K},
\end{equation}
which strongly supports the conclusion that AC-III predominantly consists of the warm neutral medium (WNM). Such a high lower limit on the spin temperature is standard for identifying a WNM-dominated component along low-optical-depth sightlines \citep[e.g.,][]{2003ApJ...586.1067H,2014ApJ...793..132S}.


\end{appendix}

\end{document}